\documentclass[useAMS,usenatbib]{mn2e}

\usepackage{amsmath}
\usepackage{amssymb}
\usepackage[parfill]{parskip}   
\usepackage{ifpdf}
\usepackage{graphicx} 
\usepackage{longtable}
\usepackage{rotating}
\usepackage{epsf}
\usepackage{bm}
\usepackage{subfigure}
\usepackage{morefloats}
\usepackage{gensymb}
\usepackage{verbatim}
\usepackage{array}
\usepackage{lscape}
\usepackage{hyperref}

\def\apjl{ApJL}

\def\apjs{ApJS}

\def\aap{A\&A}

\def\anu{1}
\def\swin{2}
\def\gemini{3}
\def\missouri{4}
\def\colby{5}
\def\harvard{6}
\def\hawaii{7}
\def\toronto{8}

\graphicspath{{figures/}}

\begin{document} 

\title[{Stellar mass - size relation} at z=1 with GeMS/GSAOI astrometry]{The {stellar mass - size relation} for cluster galaxies at z=1 with high angular resolution from the Gemini/GeMS multi-conjugate adaptive optics system.}
\author[S. Sweet et al.]
{\parbox{\textwidth}{\raggedright Sarah~M.~Sweet$^{\anu}$\thanks{Corresponding author: sarah@sarahsweet.com.au},
Robert~Sharp$^{\anu}$,
Karl~Glazebrook$^{\swin}$,
Francois~Rigaut$^{\anu}$,
Eleazar~R.~Carrasco$^{\gemini}$, 
Mark~Brodwin$^{\missouri}$,
Matthew~Bayliss$^{\colby,\harvard}$,
Brian~Stalder$^{\hawaii}$,
Roberto~Abraham$^{\toronto}$,
Peter~McGregor$^{\anu\dagger}$
}
\vspace{0.4cm}\\
\parbox{\textwidth}{$^{\anu}$ Research School of Astronomy \& Astrophysics, Australian National University, Canberra, ACT 2611, Australia\\
$^{\swin}$ Centre for Astrophysics \& Supercomputing, Swinburne University of Technology, Mail H30 P.O. Box 218, Hawthorn, VIC 3122, Australia\\
$^{\gemini}$ Gemini Observatory/AURA, Southern Operations Center, Casilla 603, La Serena, Chile\\
$^{\missouri}$ Department of Physics \& Astronomy, University of Missouri, 5110 Rockhill Road, Kansas City, MO 64110, USA\\
$^{\colby}$ Department of Physics \& Astronomy, Colby College, 5800 Mayflower Hill, Waterville, Maine 04901, USA\\
$^{\harvard}$ Department of Physics, Harvard University, 17 Oxford Street, Cambridge, MA 02138, USA\\
$^{\hawaii}$ Institute for Astronomy, University of Hawaii, 2680 Woodlawn Drive, Honolulu, HI 96822, USA\\
$^{\toronto}$ Department of Astronomy and Astrophysics, University of Toronto, 50 St George Street, Toronto, ON M5S 3H8, Canada\\
$^{\dagger}$ Deceased
}
}
\date{Released 2016 Xxxxx XX}

\pagerange{\pageref{firstpage}--\pageref{lastpage}} \pubyear{2016}

\date{\today}
 
\hypersetup{pageanchor=false}
\begin{titlepage}
\maketitle

\begin{abstract} 
We present the {stellar mass - size relation} for 49 galaxies within the $z$~=~1.067 cluster SPT-CL J0546$-$5345, with FWHM $\sim$80-120~mas $K_{\mathrm s}$-band data from the Gemini multi-conjugate adaptive optics system (GeMS/GSAOI). This is the first such measurement in a cluster environment, performed at sub-kpc resolution at rest-frame wavelengths dominated by the light of the underlying old stellar populations. The observed {stellar mass - size relation} is offset from the local relation by 0.21~dex, corresponding to a size evolution proportional to $(1+z)^{-1.25}$, consistent with the literature. The slope of the {stellar mass - size relation} $\beta$ = 0.74 $\pm$ 0.06, consistent with the local relation. The absence of slope evolution indicates that the amount of size growth is constant with stellar mass. 
This suggests that galaxies in massive clusters such as SPT-CL J0546$-$5345 grow via processes that increase the size without significant morphological interference, such as minor mergers and/or adiabatic expansion. The slope of the cluster {stellar mass - size relation} is significantly shallower if measured in $HST$/ACS imaging at wavelengths blueward of the Balmer break, similar to rest-frame UV relations at $z$~=~1 in the literature. The {stellar mass - size relation} must be measured at redder wavelengths, which are more sensitive to the old stellar population that dominates the stellar mass of the galaxies. The slope is unchanged when GeMS $K_s$-band imaging is degraded to the resolution of $K$-band HST/NICMOS resolution but dramatically affected when degraded to $K_s$-band Magellan/FourStar resolution. Such measurements must be made with AO in order to accurately characterise the sizes of {compact, $z$~=~1 galaxies}.
\end{abstract}
\end{titlepage}


\section{Introduction} 

In the local universe, most luminous galaxies belong to one of two dominant populations: early- or late-type galaxies. The former are typically passive, red, spheroids, further classified into fast and slow rotators, while the latter are generally blue, star-forming disks. 
These familiar Hubble-type classifications do not apply as readily to high-redshift galaxies, the most massive of which are compact and red \citep{Szomoru2011,Talia2014}. Few compact systems exist at the present day \citep[][]{Trujillo2009,Taylor2010,Trujillo2012,Trujillo2014}, so it is logical to suppose that the most massive high-redshift galaxies must undergo significant size evolution to become present-day passive elliptical galaxies. For example, $z\sim 1$ galaxies are $\sim$ 2 times more compact when compared with $z=0$ \citep[e.g.][]{Daddi2005,vanDokkum2008,Damjanov2009}, and those at $z\sim 4$ are $\sim$ 6 times smaller \citep{Straatman2015}. Consequently the zero-point of the {stellar mass - size relation} decreases with increasing redshift \citep[{e.g.}][]{Buitrago2008,Nagy2011,Bruce2012,Law2012,Ownsworth2014}. 

The three main proposed mechanisms for this size increase are major mergers, minor mergers and adiabatic expansion. The slope of the {stellar mass - size relation} is typically seen to be constant with redshift \citep[e.g.][]{Damjanov2011,Newman2012,vanderWel2014}, requiring that the size increase not depend on stellar mass. This disfavours major mergers, which would increase the size of the most massive galaxies at a higher rate than less massive galaxies \citep{Khochfar2006}. Major mergers are also not predicted to be sufficiently frequent to explain the observed size evolution \citep{Bundy2009,Lotz2011}. However, the situation is not clear, as some authors do observe the slope of the {stellar mass - size relation} to change with redshift, e.g. \citet{HuertasCompany2013,Ryan2012,Delaye2014}. {These studies are generally undertaken in the rest-frame UV, so size measurements may be affected by clumps of recent star formation.}

In the currently favoured minor merger paradigm described by \citet{Oser2012} and \citet{Toft2014}, compact galaxies first form out of collapsing gas, then later accrete gas-poor satellites in dry minor mergers. This is known as `two-phase galaxy formation' and was demonstrated to occur in the nearby universe by \citet{Forbes2011} with a study of globular clusters. At higher redshifts, other observational work such as \citet{vanDokkum2010,Barro2013,vanDokkum2013,Tadaki2014,vanDokkum2015} and simulations by \citet{Noguchi1999,Dekel2014,Wellons2015} have demonstrated the early formation phase, while the growth phase has been studied by \citet{Newman2012}, who found that the mergers could explain the size evolution but must happen on a rapid timescale, and by \citet{Morishita2016}, who observed sufficient numbers of satellites around a $z$~=~1.9 compact galaxy to explain the predicted size growth, once additional star formation is taken into account. In the adiabatic expansion model proposed by \citet{Fan2008,Fan2010}, galaxies experience a rapid mass loss event caused by AGN or supernova winds. After some time delay, expansion occurs in an amount proportional to the fraction of mass lost \citep{RangoneFigueroa2011}. Recent analysis by \citet{Wellons2016} of galaxies in the Illustris simulation \citep{Genel2014,Vogelsberger2014a,Vogelsberger2014b,Nelson2015} suggests that adiabatic expansion is responsible for less size growth than mergers are. Adiabatic expansion associated with the formation of new stellar populations {(but not due to AGN winds) during 2~$> z >$~1.2} was essentially ruled out by \citet{Damjanov2009}{, though that study was conducted at resolutions similar to the angular sizes of galaxies at those redshifts}. 

The {stellar mass - size relation} can be used to distinguish between major mergers and minor mergers / adiabatic expansion. 
{The accuracy of the relation depends most strongly on resolution and rest-frame wavelength.\footnote{Limiting magnitude appears less critical than spatial resolution for morphological classification; \citep[][]{Povic2015}. For cluster galaxies, sufficient spatial coverage of the cluster is also required in order to remove any environmental effect \citep[e.g.][]{Strazzullo2010}.} On one hand, s}ufficient resolution is required to measure effective radii and S{\'e}rsic \citep{Sersic1968} indices of the most compact galaxies. {On the other hand, and equally i}mportantly, the rest frame redwards of the 4000 \AA\/ spectral break is necessary for this measurement, as it is stellar mass that is the main driver of galaxy properties such as colour, age and specific star formation rate. Current efforts to measure high-redshift galaxy morphologies at high resolution {\citep[e.g.][]{HuertasCompany2013,Ryan2012,Delaye2014}} are limited to optical HST imaging \citep[e.g. FWHM 0.09 arcseconds in F814W $\sim I$-band,][]{Scoville2007}, but such filters are rest-frame UV at these redshifts. Consequently, they suffer from contamination of starburst events rather than tracing the bulk of the stellar mass. Near infrared imaging has necessarily poorer resolution, e.g. HST/WFC3 $\sim H$-band FWHM is 0.18 arcseconds \citep{Guo2011}, which is 1.48 kpc at $z=1$ and thus may not provide sufficient resolution for accurate profile fitting {\citep[e.g.][]{Damjanov2009}}. Due to this tradeoff between resolution and wavelength, there is a wide variety of rest-frame wavelengths at which the measurements are made, with correspondingly wide variety in the results. \citet{Carrasco2010} addressed this tradeoff with ground-based adaptive optics imaging, but were limited to just eight field galaxies observed one at a time due to the small effective field of view of the Gemini NIRI camera when corrected by the Altair laser guide star system. {Consequently, the current literature cannot be used to distinguish between major mergers, minor mergers and adiabatic expansion.} 

In this paper we study the massive, evolved galaxy cluster SPT-CL J0546$-$5345. We present the first high-resolution ground-based measurement of the {stellar mass - size relation} measured in the rest-frame light of the old stellar population at $z$~=~1, and demonstrate the importance of measuring the relation redward of the 4000 \AA\/ break by comparing with rest-frame UV {stellar mass - size relation}s for the same sample. In Section 2 we present our GeMS $K_s$-band imaging and describe our observing and data processing strategies, including successful correction of the optical distortion and faint source residual images. We also present our ancillary data of GMOS spectroscopy and HST archival F606W and F814 band imaging. Section 3 contains the galaxy profile fitting and cluster membership.
The {stellar mass - size relation} is presented and discussed in Section 4. In Section 5 we summarise our conclusions.

Throughout this work we assume a concordance cosmology with $\Omega_M$=0.264, $\Omega_\Lambda$=0.736 and H$_{0}$=71 km s$^{-1}$ Mpc$^{-1}$, for consistency with \citet{Brodwin2010}.  Magnitudes are presented on the AB system. 

\section{Galaxy cluster selection and observations - SPT-CL J0546$-$5345}

The Gemini MCAO (multi-conjugate adaptive optics) system \citep[GeMS;][]{Rigaut2014,Neichel2014a} mounted on the 8.1-m Gemini South telescope uses five laser guide stars \citep{d'Orgeville2012}, and up to three natural guide stars, to perform a multi-conjugate adaptive optics correction across the $\sim$85$\times$85\arcsec\ field of view of the Gemini South Adaptive Optics Imager \citep[GSAOI,][]{McGregor2004,Carrasco2012}. The natural guide stars are required to correct for the tip-tilt atmospheric component, which is not sensed by the laser guide star system. The number of high redshift galaxy clusters accessible to GeMS is limited by the magnitude restrictions ($R \le$15.5 mag) for the first generation natural guide star wavefront sensors. Within these constraints we identified our first target cluster as the $z$~=~1.067 structure SPT-CL J0546$-$5345 \citep{Brodwin2010} detected as part of the 2500 square degree South Pole Telescope Sunyaev-Zel'dovich survey \citep[][]{Bleem2015b}. 
The cluster has a virial mass of M$_{200} = 1.0^{+0.6}_{-0.4}$ x 10$^{15}$ M$_\odot$; although it is massive for its redshift its existence is not in conflict with $\Lambda$CDM \citep{Brodwin2010}.  
At the redshift of the cluster 1$\arcsec$ = 8.2 kpc. {The virial radius is $r_{200}$ = 1.57 Mpc = 191\arcsec \citep{Brodwin2010}.}

\begin{figure*}
\centerline{
\includegraphics[width=1\linewidth]{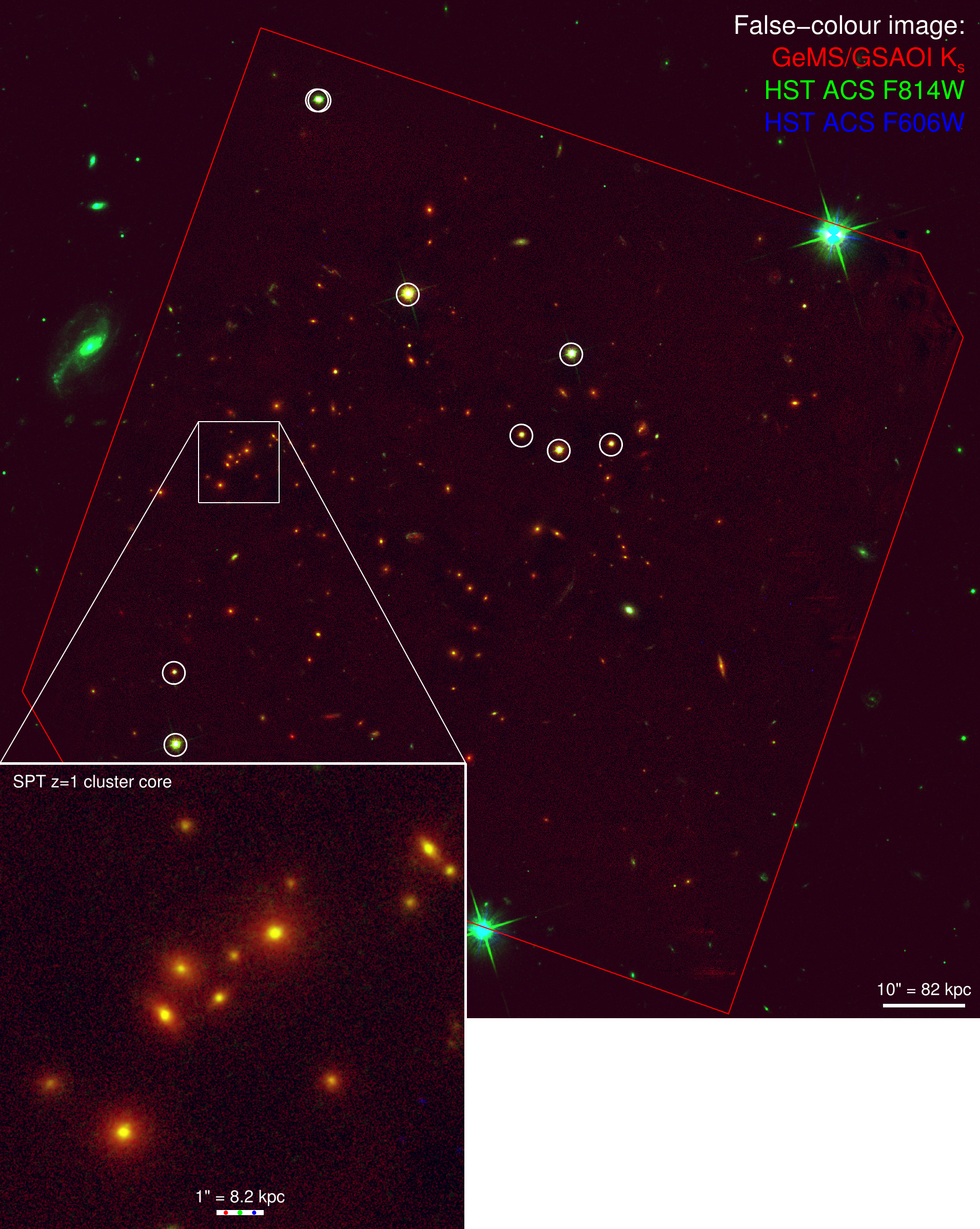}
}
\caption{\label{sci} False-colour image of SPT-CL J0546$-$5345 at $z=1.067$. Gemini GeMS/GSAOI Ks (this work) = red, HST ACS F814W = green, HST ACS F606W = blue. The red polygon shows the approximate sky coverage of the GSAOI pointings. PSF stars are indicated by white circles. {The star near the top of the image with two circles is a binary star.} The inset at lower left is a zoom-in of the cluster core. The scale bars at lower right of the inset and full image show the angular and physical projected distances at the cluster redshift. The red, green and blue spots on the scale bar in the inset show the PSF FWHM for each band. N is up and E is left.
}
\end{figure*}

\begin{figure*}
\centerline{
\includegraphics[width=1\linewidth]{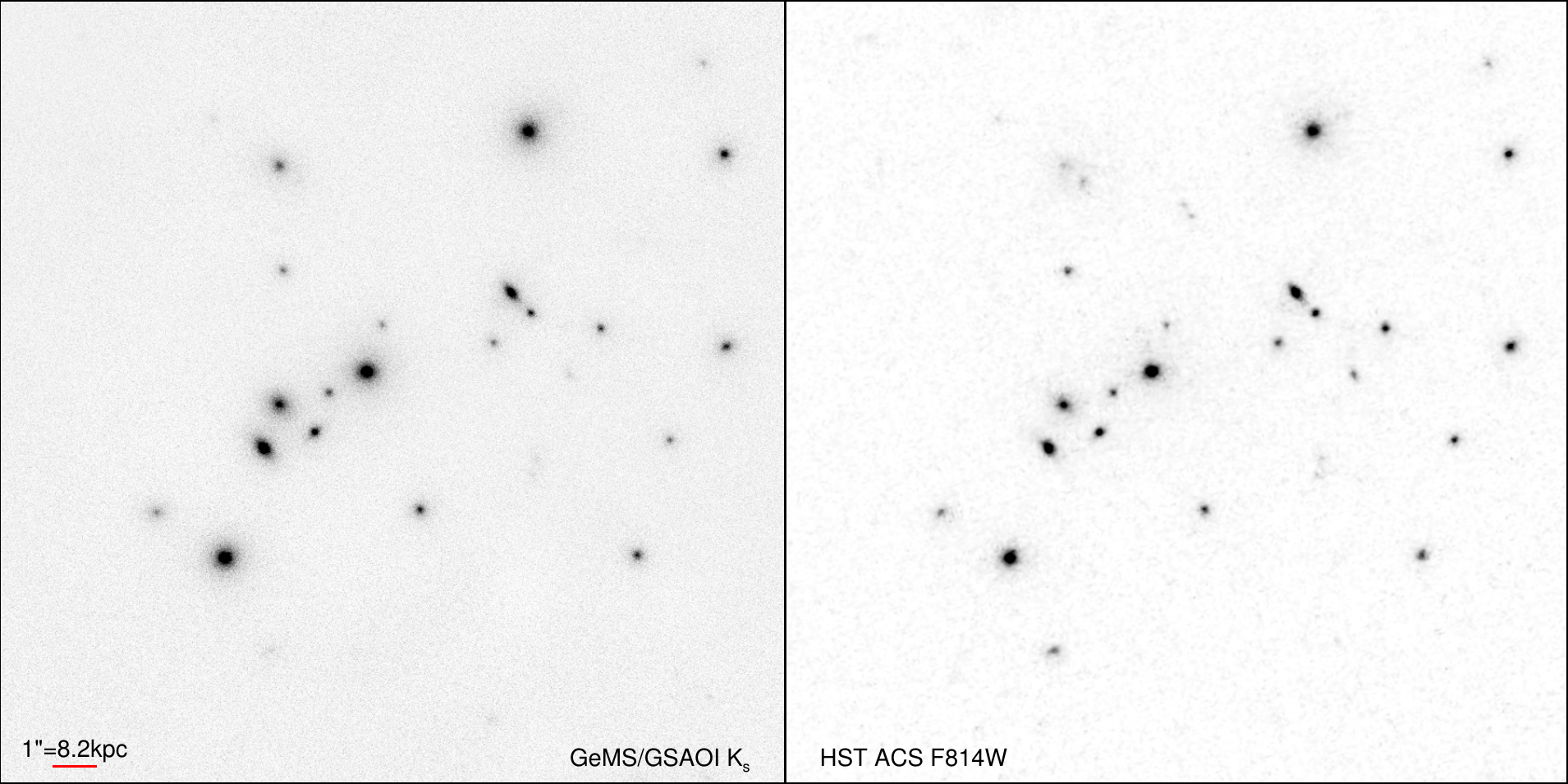}
}
\caption{\label{core} Zoom-in of cluster core. Left: Gemini GeMS/GSAOI K$_s$; Right: HST ACS F814W. The scale bar at lower left shows the angular and physical projected distances at the cluster redshift. 
N is up and E is left.
}
\end{figure*}

In the next subsection we present our GeMS $K_{\mathrm s}$-band imaging of this cluster. We also retrieve archival $HST$/ACS F606W and F814W imaging of the cluster of comparable depth from program 12477 (PI: High); this is presented in \S~\ref{hstdata}. (See Figure~\ref{sci} for a three-colour image of the GSAOI field, and Figure~\ref{core} for a comparison between GeMS/GSAOI K$_s$ and HST ACS F814W.) In \S~\ref{gmosdata} we present new Gemini/GMOS spectroscopy of sources within the GeMS field of view.

\subsection{Near-infrared wide field AO imaging observations}
\label{gemsdata}
The GeMS near-infrared camera, GSAOI, is a 2$\times$2 mosaic of four Teledyne HAWAII-2RG HgCaTe detectors, with an inter-chip gap of $\sim$2.7\arcsec. The nominal pixel scale of GSAOI is 19.7 mas, which {samples the $J$-band Nyquist} detection limit, and somewhat oversamples\footnote{Since observations are background-limited in modest duration exposure with broadband filters, this oversampling has little penalty.} the $K_s$-band PSF. Operating at the diffraction limit of the 8.1-m Gemini telescope, GeMS/GSAOI achieves 2.25 times better angular resolution (in the $H$-band) than $HST$/WFC3 $H_{160}$-band, which has a PSF FWHM 0.18$\pm$0.01\arcsec\ sampled with 0.08\arcsec\ drizzled pixels \citep[e.g.][]{Law2012}.

Cluster SPT-CL J0546$-$5345 was observed with GeMS in guaranteed time observations during 2013 November 19-21 (Program GS-2013B-C-1) and 2014 December 05-08 (Program GS-2014B-C-1), using the two natural guide stars accessible to the wavefront sensor. We conducted preliminary tests on our relatively shallow 2013 dataset before gathering deeper observations in 2014. The later observations are presented in this work. Natural seeing FWHM ranged from 0\arcsec.6 to 1\arcsec.2.

There are static distortions present in the GeMS/GSAOI system, caused by optical design and physical alignment of the four arrays. We account for these via careful measurement of standard astrometric fields (with a high star density of tens per detector, and good relative astrometry) taken at the same position angle as our science data during the observing block. Previous work had warned of small but significant changes to the variable plate scale solution when observing with large ($>3$\arcsec) relative offsets, due to the design of the Canopus AO system wavefront sensor feed arrangement \citep{Rigaut2014}; however, our data show this concern to be unfounded. 
To fill in the inter-chip gaps and provide a contiguous field of view in the final composite image, we conducted dithered observations offset by [4,4], [0,0], [-4,-4]~\arcsec\ relative to our base pointing.
To maximise the rejection of cosmetic defects on the detectors, and to minimise correlated noise artefacts when generating sky frames, we also conducted nine micro-dithers with a pitch of $\sim$ 0\farcs3 (i.e., by more than the angular resolution element) at each of the three large offset positions.
These micro-dithers were small enough to allow image stacking without prior distortion correction. The advantage of this method over a random dither pattern is that it allowed sufficient signal-to-noise ratio detections of faint field sources within each base position observation to facilitate a secondary, relative offset correction between the large dither positions. We interleaved the large and micro-dithers, yielding an observation sequence that might be described as ABCA{\arcmin}B{\arcmin}C{\arcmin}A{\arcsec}B{\arcsec}C{\arcsec}...~.  In this way, the frames at each large dither position (e.g. A) have two alternative large dither positions (B and C), whose micro-dithers (\arcmin,\arcsec,...) serve as sky frames. Such a sequence of three large dither positions by nine micro-dither observations of 120~sec duration completes an on-source exposure sequence of 27 independent frames. The resulting on-source exposure time of 0.9~hours requires an elapsed time of 1.5~hours, after accounting for telescope offsetting and near-infrared array readout time; that is, an efficiency of 60\%. Longer individual observations were considered to be at risk from increased sky variability (although we find limited evidence for this throughout the course of our runs) while shorter integrations would deliver a lower efficiency in the duty cycle. This 27-exposure pattern was executed at six different base positions for a total of approximately 5.5 hours on source. 

\subsubsection{Data reduction}
The basic data reduction steps were conducted with the Gemini IRAF GSAOI package\footnote{The Gemini IRAF software is available at http://www.gemini.edu/sciops/data-and-results/processing-software.}, with data from individual detectors stored as separate extensions within multi-extension fits files. Daytime calibration {\it lamp-on} and {\it lamp-off} dome flats were combined to generate a master dome flat. Variance and data quality extensions were populated and non-linear or saturated pixels were flagged. The frames were flat fielded using dome flats, as we found that these gave lower residual structure compared to twilight or dome {multiplied by} twilight sky flats.

\subsubsection{Spatial distortion correction and mosaic alignment}
The GeMS imaging system introduces a $\sim$3\% variation in the plate scale across the field of view \citep{Neichel2014b,Schirmer2015}.
The image distortion is largely static, with small, variable distortions that depend on instrument flexure under gravity and the positions and relative magnitudes of the guide stars.
When there are many bright sources within the field of view, this distortion may be corrected by referencing to those sources. For example, \citet{Schirmer2015} report $K_s$-band residuals of 10~mas (internal) and 13~mas (with respect to external astrometry). Much of the error for that correction arose from centroiding to galaxies rather than stars. The resulting PSF in their stacked image was 70$<$FWHM$<$100~mas across the field, with the best resolution being nearest the single natural guide star used during observations.

When there are few bright sources within the field of view, as is generally the case for faint galaxy clusters at high redshift, a separate astrometric field must be observed. The astrometric field must have a deep reference catalogue with similar resolution to GSAOI.
The chosen field should be observed during the same run as the science data, and have the same {position angle (PA)} and, where possible, airmass. Ideally, the positions and relative magnitudes of the guide stars should be matched to those in the science frames. 
An astrometric distortion solution was derived for this work from contemporary observations of a field located within globular cluster NGC\,288 \citep[catalogued in][]{Anderson2008}. Six 30-s exposures and five 30-s {exposures offset by $\sim 6\arcmin$ to serve as sky frames} were taken at the same {PA} as our science observations, using two natural guide stars in a similar asterism and with similar relative magnitudes to those in the science field. Basic reduction was performed with the GSAOI IRAF package. We used the GAIA software package \citep{Draper2014} to perform a coarse, initial {world coordinate solution (WCS)} solution for each detector in the GSAOI mosaic. Following that we constructed a catalogue of sources in the astrometric field using SExtractor \citep{Bertin1996} and compared this to the HST catalogue \citep{Anderson2008} with SCAMP \citep{Bertin2006}. We applied the resulting distortion correction to our science images with SWarp \citep{Bertin2002}, which mosaicked the individual detectors to a single image. {We provide more detailed instructions at Appendix~\ref{distortion}.} The mosaicked images were reregistered to a common reference by centroiding the single bright star common to our science images. The astrometric rms is 5 mas internally and 8 mas with respect to HST ACS F814W imaging. 
The final combined image PSF FWHM varies across the field from 80\,mas close to the centre of the field to 130\,mas at the edges of the field of view; measured from eight stars in the image (see Appendix~\ref{psfchar}). The maximum Strehl ratio is 10\% (median 8\%), consistent with the predictions of the Gemini Observing Tool, which takes into account the number, position and magnitude of the natural guide stars. The median 50\% and 85\% encircled energy diameters are 0.22\arcsec and 0.50\arcsec respectively. 

\subsubsection{Sky subtraction residual image removal}
\label{skysub}
\begin{figure}
\centerline{
\includegraphics[width=1\linewidth]{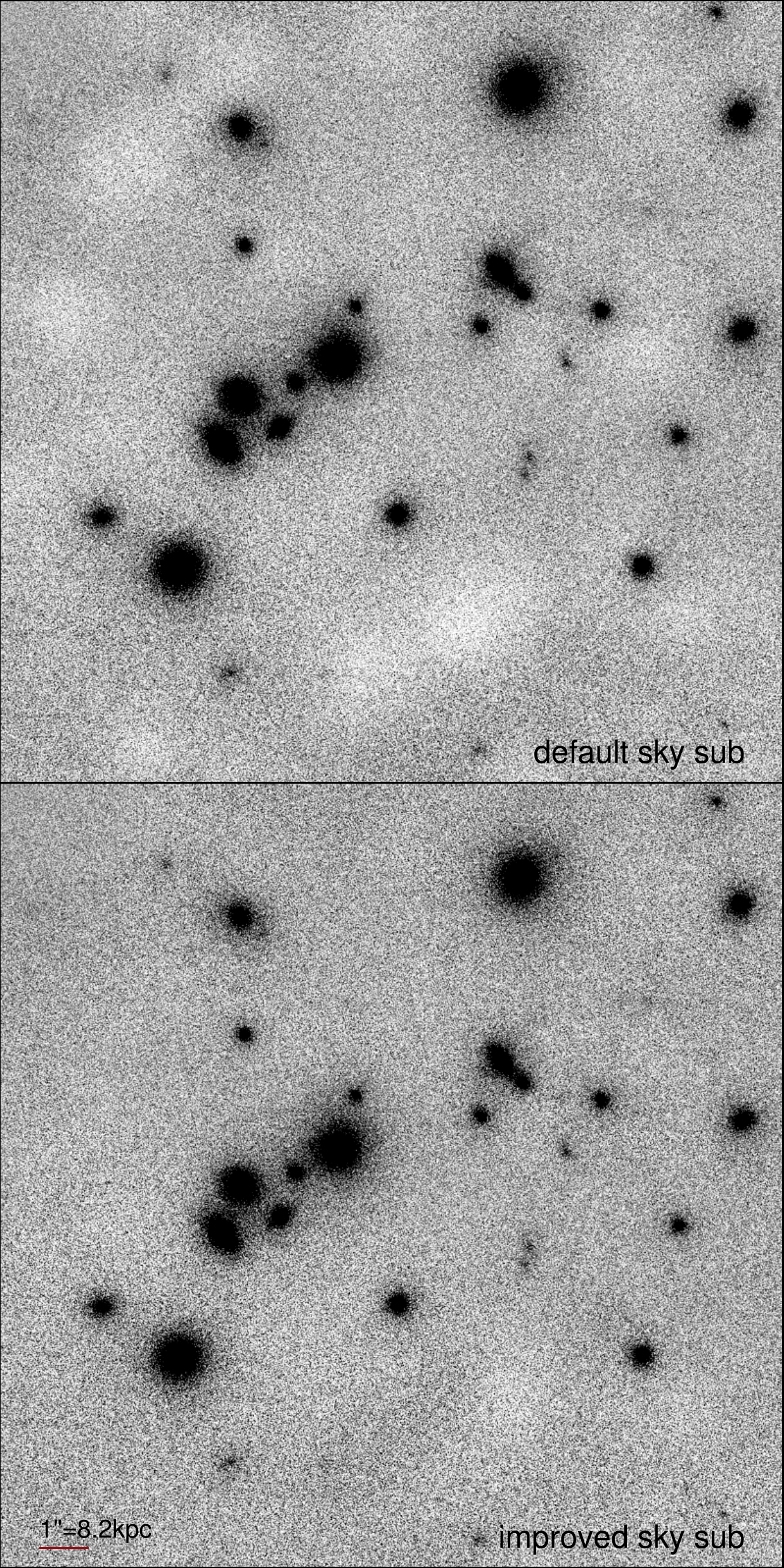}
} 
\caption{\label{residual}Example section of GeMS image showing white residuals appearing in default sky subtraction (top) and reduced by the sky subtraction method outlined here (bottom). The scale bar at lower left shows the angular and physical projected distance at the cluster redshift. 
}
\end{figure}

The default median sky subtraction method results in significant subtraction residuals on the order of 1-2 times sky rms, coincident with the large dither pattern positions \citep[see][and our Figure~\ref{residual}]{Schirmer2015}. This is a consequence of using a regular, semi-repeated dither pattern with offset object frames used as sky frames and insufficient masking of faint sources.

A larger pseudo-random dither pitch may alleviate this problem; however, early concern over differential distortions between larger offset dithers, and the lack of a larger number of independent restoration stars in our field, led to the observing strategy described above.

The faint sources causing the residual images are not detected in any individual 120-s image frames and therefore cannot be masked when the sky frames are constructed. When combining $\sim$ 200 frames for the final stacked image the resulting residuals become significant at the level of the typical rms sky noise. We overcame this issue by creating an improved object mask from the stacked image. We calculated the inverse distortion correction for each detector, and distorted the object mask back to the observational reference frame of each detector in each input observation. {Detailed instructions are provided at Appendix~\ref{apresidual}.} As demonstrated in Figure~\ref{residual}, the residuals are significantly reduced. The remaining variation is accounted for in our Monte Carlo error estimation (\S~\ref{error}).

{Sky frames were created for each large dither position by combining the 18 frames from the alternate large dither positions. 
These sky frames spanned the 1.5 hours of a single exposure sequence.  
Experimentation with sky frames spanning shorter time intervals showed increased sky noise and consequently poorer sky subtraction. Each science frame was then sky subtracted using a median scaling of the relevant sky frame. }
We then returned to the GSAOI IRAF package at the sky subtraction step, performing the distortion correction as before.

\subsection{Image stacking and flux calibration}
\label{flux}
We discarded the 15\% of frames that have average PSF FWHM $>$ 6 pixels (120 mas) before combining the available data for the final stacked image, for a total exposure time of 4 hours 42 min. Individual frames were median scaled to multipliers computed from relative source fluxes, in order to account for differences in airmass between the individual frames.

The final combined image was flux calibrated by comparing measurements for the bright star in the field from GSAOI and 2MASS \citep{Skrutskie2006}. {The 2MASS K$_s$ measurement for this star is 15.508 $\pm$ 0.188 Vega mag. We assumed} a conversion correction between the Vega and AB magnitude systems of m{$_{Ks}$}(AB)= m{$_{Ks}$}(Vega) + 1.86 mag \footnote{https://www.gemini.edu/sciops/instruments/gsaoi/calibrations/photometric-standards-and-zero-points}.

The limiting surface brightness rms per pixel at 3$\sigma$ above sky level is $\mu_{K_s} $= 16.22 {$\pm$ 0.19} (AB) mag~/~arcsec$^2$.

\subsection{Short wavelength imaging observations}
\label{hstdata}
Archival high angular-resolution imaging data is available for SPT-CL J0546$-$5345, with $HST$/ACS imaging in the F606W and F814W broad-band filters observed as part of program 12477 (PI: High). We obtained multi-drizzled $HST$/ACS data from the Mikulski Archive for Space Telescopes\footnote{http://archive.stsci.edu/}. Exposure times were 1920 s for F606W and 1916 s for F814W. Zero points were calculated in a similar manner to \S~\ref{flux}.

A three-colour composite image of SPT-CL J0546$-$5345 is shown in Figure~\ref{sci}. The native resolution of the $HST$/ACS F814W ($\sim I$-band) image is equivalent to that of the AO-assisted Gemini/GeMS $K_{s}$ image, as is highlighted by the close-up of the cluster core region shown in 
Figure~\ref{core}.

\subsection{GMOS spectroscopy for cluster membership confirmation}
\label{gmosdata}
In order to augment the number of known cluster members for SPT-CL J0546$-$5345 \citep[Table~\ref{cat};][]{Brodwin2010,Ruel2014} we performed additional follow-up spectroscopy with the GMOS spectrograph at Gemini-South (Program GS-2013B-Q-70). 
Observations were performed with the GMOS-S EEV detectors, which have since been replaced with the more sensitive Hamamatsu array. The R400+G5325 grating+filter combination was centred at a wavelength of 8200~\AA\ to target the [O\textsc{ii}]\,$\lambda$3727 emission line (when present), the 4000\AA\ spectral break and the Ca\,H+K absorption features in galaxies at the cluster redshift. Unfortunately the H$\alpha$ emission-line, prominent  in AGN or star-forming galaxies, lies at an observed frame wavelength of $\lambda$~=~1.3565\,$\mu$m for sources at the cluster redshift of $z$~=~1.067, placing the line within the region of strong atmospheric water absorption between the near-infrared $J$ and $H$-bands. Consequently, near-infrared H$\alpha$ spectroscopy, which is valuable as a tracer of ongoing star formation as well as a convenient redshift indicator when present, is not practical at this redshift with ground-based MOS facilities.

Due to the limited angular extent and high candidate source density of the compact cluster region, we performed the observations in the {\it nod-and-shuffle} {(N\&S)} observing mode \citep{Glazebrook2001}, using {\it band shuffling} \citep{Abraham2004}.
This observation mode not only allows high-quality sky subtraction but also allows short slits such as the 1\arcsec\ apertures used in this work. In this way a high slit density can be achieved in the central third of the 1-arcminute GMOS field, which corresponds roughly to the 85\arcsec\ field size of the GeMS imaging.
Two slit masks were observed, each for 3$\times$1800~s at three central wavelengths (8200~$\pm$~100~\AA) to fill in the wavelength gaps introduced in the spectral range by the gaps in the CCD mosaic. This resulted in a total integration time on each mask of 4.5~hours.
At the time of mask preparation neither the GeMS imaging (\S\ref{gemsdata}) nor the $HST$ imaging (\S\ref{hstdata}) was available, so the target catalogue was comprised of detections in the GMOS $I$-band \emph{preimage}. Target galaxies were prioritised in the following order:
\begin{enumerate}
\item likely cluster members within the proposed GeMS imaging field of view, particularly sources in the densely-populated region close to the assumed brightest cluster galaxy;
\item similar sources within the GMOS mask field of view but outside the GeMS imaging footprint;
\item repeat observation of known cluster members, to provide a convenient cross-check of the new observations; and,
\item filler targets including a tentatively detected gravitational lens arc, potential faint PSF stars and probable foreground galaxies.
\end{enumerate}
The observations were processed using the standard Gemini IRAF packages. 
The {56} reduced spectra {are uniformly distributed across the GMOS field of view; 33 fall within the smaller GSAOI pointing. Redshifts for these were measured} manually using the \textsc{runz} template cross-correlation software (developed originally by Will Sutherland for the 2-degree Field Galaxy Redshift survey \citep{Colless1999,Colless2001}). The resulting identifications of six new cluster members and two foreground galaxies are included in Table~\ref{cat}.

\section{Cluster membership}
We measured sources in the three imaging bands with SExtractor \citep{Bertin1996}. We constructed a catalogue of sources that are common to all three imaging bands and have $K_s$ and F814W signal-to-noise ratio $>$ 4, and signal-to-noise ratio $>$ 1 in the shallower F606W band\footnote{{The low signal-to-noise threshold in the F606W band does not adversely impact our work. Visual inspection of the model fits shows that all but two of the galaxies (45 and 48) are well fit in F606W; those two are excluded from our analysis in that band. }}. We define cluster members as those galaxies that either are spectroscopically confirmed to lie at the cluster redshift, or have similar photometric properties, as per the following subsections. {Sources in the field are presented in Table~\ref{cat}, with cluster member properties given in Table~\ref{galprop}.}

\subsection{Spectroscopic redshifts}
Seventeen sources in the observed field are known to be cluster members on the basis of ground-based spectroscopy. Of the 124 sources identified within the GeMS $K_s$ field, five are previously-confirmed cluster member galaxies with spectra published in \citet{Brodwin2010} and \citet{Ruel2014}. One of these (our ID 50) has different redshifts given in the two catalogues; the rest have consistent redshifts in both papers. A further five spectra have redshifts 0.9 $<z<$ 1.15 in our GMOS N\&S spectroscopy. Two of these spectra each contain the flux from two objects (IDs 69 \& 72, and 85 \& 89) identified in the high angular resolution imaging. We can only detect one redshift in each spectrum. In the case of 69 \& 72 it is not clear to which object the redshift belongs. However, since the blended sources have similar colours and visual morphologies, and their projected location places them within the densely populated core region of the cluster, we assign cluster membership to both sources. In the case of 85 \& 89, the two sources have distinct colours and morphologies, so we assign membership to the brighter source 85. Our GMOS spectroscopy thus yields six new cluster members. Their spectra are generally quiescent. We also detect two foreground galaxies at $z$~=~0.4 and 0.5.

The cluster is being observed by the GOGREEN\footnote{gogreensurvey.ca} survey (PI: Michael Balogh), who have provided preliminary spectroscopic redshifts for seven cluster members within our GSAOI field of view. One of these is common to our GMOS spectroscopy {(its GOGREEN redshift is within $\Delta z$ = 0.03), while} the remaining six are newly-identified cluster members, bringing the total to 17 spectroscopically-confirmed cluster members in our field. These have a median redshift $z$~=~1.0669 $\pm$ 0.009.

We note that outside of the GSAOI field of view, there are additional GMOS spectra as well as more \citet{Brodwin2010} and \citet{Ruel2014} members, but guide star limitations prevented us from observing them with GeMS/GSAOI {so they are not studied in this work}. Three of these galaxies both were observed with GMOS N\&S and appear in the \citet{Brodwin2010} and \citet{Ruel2014} catalogues. For these three sources our GMOS redshifts were measured independently of, and are {on average within $\overline{\Delta z}$ = 0.001 of}, the previously published redshifts.

\subsection{Photometrically-selected sources}
\begin{figure*}
\centerline{
\includegraphics[width=1\linewidth]{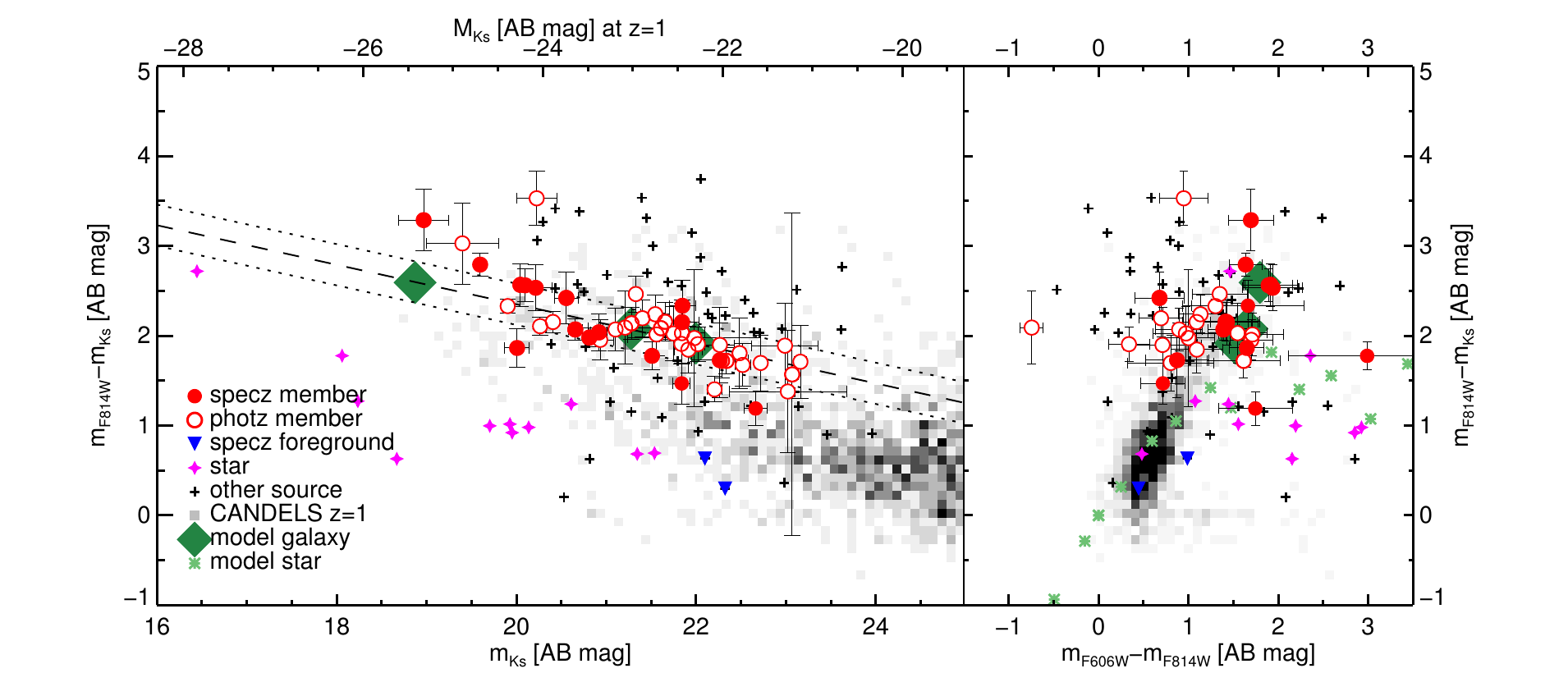}
}
\caption{\label{cmd}  {\it Left:} Colour-magnitude diagram demonstrating photometric selection of additional cluster members. Red filled circles denote spectroscopically-confirmed cluster members. 
The grey density field depicts $z$~=~1 galaxies in CANDELS \citep{Guo2013}, and the dashed line shows the line of best fit to the red sequence defined by that sample. Red open circles are photometric members either selected by eye as likely cluster members based on colour, size and morphology, or within $\pm$3$\sigma$ (dotted lines) from the red sequence. Blue triangles are spectroscopically-confirmed foreground galaxies. Magenta stars are known stars with obvious diffraction spikes in the HST imaging or similar photometric properties. Black plus symbols show other sources in the field of view. Large, dark green diamonds depict model galaxies generated using a \citet{Bruzual2003} single burst model normalised to the CANDELS sample, with metallicities Z = 0.05, 0.02 (solar) and 0.008.
{\it Right:}  Colour-colour diagram demonstrating validity of star colours. Green asterisk symbols are median binned model stars in the stellar libraries of \citet{Lejeune1997} with [M/H] = $-$1.50. The model stars show reasonable agreement with our measured stars in the field of view (magenta star symbols). For clarity we only show sources with signal-to-noise ratio $>3$ in F606W, though the less stringent cut is used in our analysis.
}
\end{figure*}

{A further 32 cluster members are identified by colour selection, for a total of 49 members. 
They} were either not targeted for spectroscopy (due to the limitations of slit packing with GMOS, even with the use of the nod-and-shuffle technique), or their GMOS spectra were of insufficient signal-to-noise ratio to yield a redshift. In the left-hand panel of Figure~\ref{cmd} we show a colour-magnitude diagram for all sources in the field of view. The spectroscopically-confirmed cluster members are plotted with red, filled circles. 
The grey-scale density plot shows CANDELS sources at $z=1$, with photometry from \citet{Guo2013} and photometric redshifts from \citet{Hsu2014}. We show the best-fitting line to the CANDELS red sequence, with the dotted lines showing the $\pm$3$\sigma$ range. 
{Sources within the 85\arcsec\ field that satisfy $m_{F814W} - m_{K_s} \le 6.98 - 0.22m_{K_s}$ and $m_{F814W} - m_{K_s} \ge 6.52 - 0.22m_{K_s}$ are within that range and defined to be be photometrically-selected cluster members. There are 34 sources selected in this way, of which 27 (55\% of the total sample of 49 members) do not have spectroscopic redshifts.} 
We add to that sample {a further 5 (10\%)} sources outside of that range with similar colour, size, morphology and location to the confirmed cluster members. Other sources red-ward of that range are likely $z>1$ red galaxies. 

Also shown on Figure~\ref{cmd} are \citet{Bruzual2003} single burst model galaxies of metallicities Z = 0.05, 0.02 (solar) and 0.008 formed at $z_f=3.0$, generated with EzGal\footnote{www.baryons.org} \citep{Mancone2012} and normalised to the CANDELS red sequence. 

In the right-hand panel of Figure~\ref{cmd} we plot a colour-colour diagram for stars in the GSAOI field. We use the stellar population synthesis model of \citet{Robin2003}\footnote{model.obs-besancon.fr} to simulate Milky Way stars in the direction of our cluster, finding a median  metallicity [Fe/H] $\sim -$1.7. We then choose the model stars with closest-matching metallicity ([M/H] = $-$1.50) from the \citet{Lejeune1997} stellar libraries. For clarity we bin in optical colour and plot the median infra-red colour in each bin. There is a reasonable agreement between those model stars and the observed stars in our field of view.

\subsection{Stellar mass estimates}
In order to robustly estimate stellar masses we supplement our high resolution imaging with follow-up SPT survey imaging presented in \citet{Song2012}, namely $riz$ MosaicII photometry from the Blanco Cosmology Survey \citep{Desai2012,Bleem2015a} and 3.6 and 4.5 $\mu$m Spitzer IRAC photometry (PI Brodwin; program ID 60099).

We use the FAST \citep[Fitting and assessment of synthetic templates][]{Kriek2009} code to fit \citet{Bruzual2003} models to the photometry, using a \citet{Chabrier2003} IMF and \citet{Calzetti2000} dust law. Our grid of exponentially declining star formation history models $SFH \sim {\rm exp}(-t/\tau)$ includes timescales 6.5 $\leqslant$ log($\tau$) $\leqslant$ 11 in steps of 0.5, metallicities $z$~=~0.008, 0.02 (solar) and 0.05, and dust extinction 0~$\leqslant A_v \leqslant$~3. We limit ages to older than 10 Gyr and younger than the age of the Universe (that is, formation epochs prior to $z=$ 2).

Resulting stellar masses are given in Table~\ref{galprop}. We note that FAST gives similar (0.2 dex lower at 10$^{11}$~M$_\odot$) masses to those obtained with a simple $K$-band mass-to-light ratio conversion at $z$~=~1.1 derived by \citet{Drory2004} and given in Table~1 of that paper. 

\subsection{Galaxy profile fitting with GALFIT}
\begin{figure}
\centerline{
\includegraphics[width=1\linewidth]{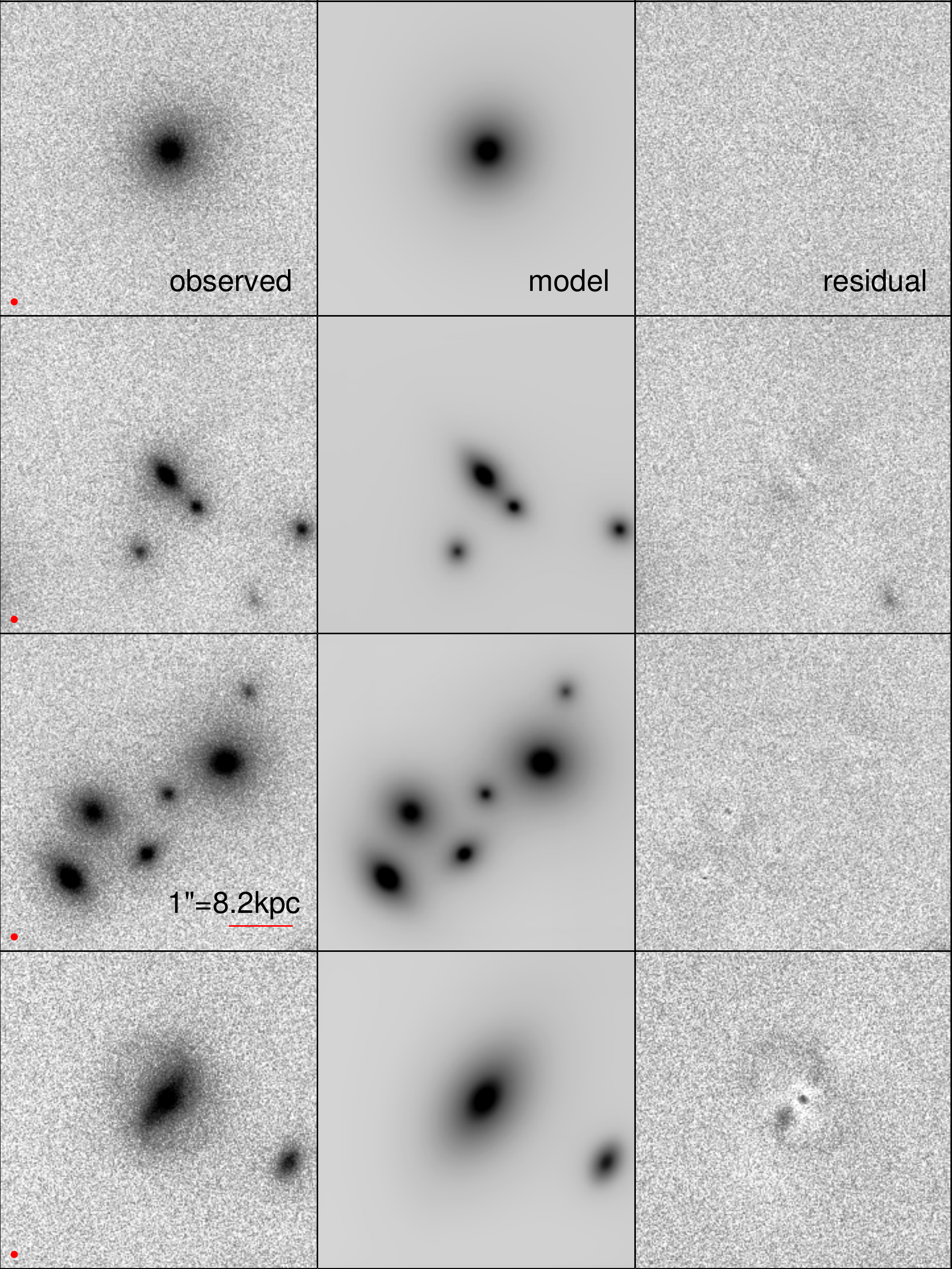}
}
\caption{\label{galfits} Profile fitting of a selection of sources in $K_s$-band using a GALFIT single S{\'e}rsic profile. Images are scaled to a common inverse hyperbolic sin scaling to emphasise faint residual features. Left to right: observed, S{\'e}rsic fit, residual. Top to bottom: isolated, blended, cluster core, foreground spiral galaxy. The red spot in the lower left corner of the observed panels illustrates the average PSF FHWM. The scale bar near cluster core shows the angular and projected distance at the cluster redshift of z=1. N=up and E=left in each panel.
}
\end{figure}

We fit the cluster member galaxies using the IRAF interface to GALFIT \citep{Peng2002,Peng2010}. We use our interpolated Moffat profile PSF described {in detail in Appendix~\ref{psfchar}} for the GeMS $K_s$ imaging. For the $HST$ data we choose not to adopt the common methodology of creating a PSF using the Tiny Tim \citep{Krist2011} software, which is ideal when the PSF is under-sampled, but does not model the PSF of {\it multi-drizzled }data. In our case, the PSF is well-sampled due to the multi-drizzling process, and consistent across the field of view, so we use an isolated star PSF for profile fitting. Considering the modest signal-to-noise ratio and angular extent of the sources of interest in the SPT-CL J0546$-$5345 cluster we adopt a single S{\'e}rsic profile fit to each source. A selection of profile fits are shown in Figure~\ref{galfits}, illustrating the validity of the method for isolated, blended and cluster core galaxies. 
{There are no obvious mergers in the sample.}

\begin{figure}
\centerline{
\includegraphics[width=1\linewidth]{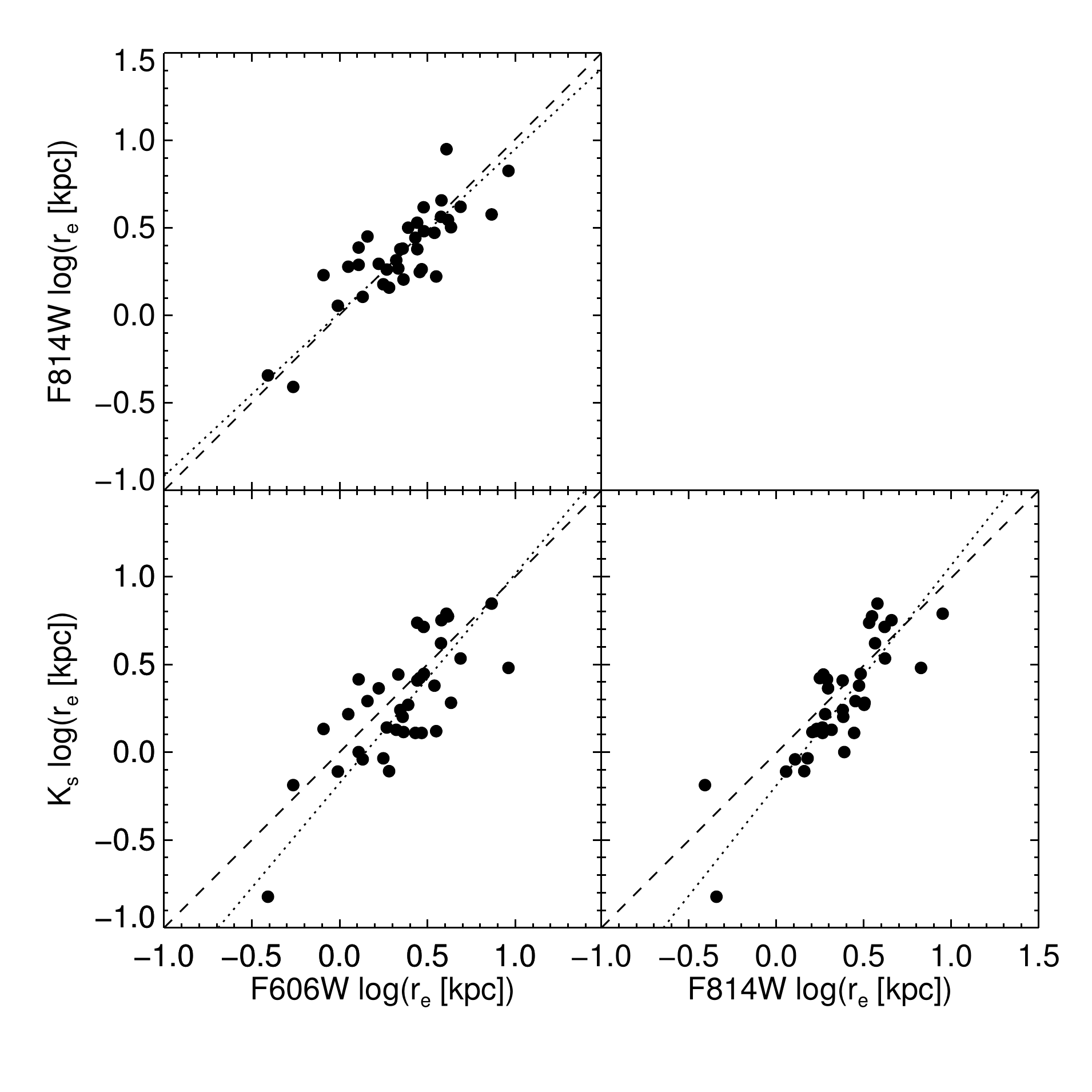}
}
\caption{\label{comparefits} Cluster member effective radii measured in HST F606W-band vs. HST F814W-band (top); $K_s$-band vs. HST F814W-band (bottom right); $K_s$-band vs. HST F606W-band (bottom left). The dashed line in each panel is the 1:1 relation; the dotted line is the best fitting line. Spearman's rank correlation coefficients are 0.756, 0.834 and 0.712 respectively.
}
\end{figure}

For consistency with the literature we measure circularised effective radius $r_{e} = \sqrt{ab}$, where $a$ and $b$ are the effective semi-major and -minor axes fitted by GALFIT\footnote{The output of GALFIT contains $a$ and $q=b/a$.}. 
We compare galaxy effective radii between the three bands in Figure~\ref{comparefits}. Visually, measurements are not well matched between pairs of bands, a natural consequence of variation in the stellar populations probed at different wavelengths. We point out that measured resolutions of the F606W and F814W images are similar to our mean $K_s$-band image resolution (111, 132 and 112 mas respectively), so differences in resolution are not responsible for the differences in measured $r_e$; we have isolated the effect of rest-frame wavelength. 
We find that there is a clear trend with a large amount of scatter in each case, reflected in the moderate Spearman's rank correlation values of 0.756 for F606W vs. F814, 0.834 for F814W vs. $K_s$ and 0.712 for F606W vs. $K_s$. The scatter largely results from the $HST$ imaging tracing the rest-frame UV, and the size measurements are therefore biased by clumpy star formation regions. Indeed, for some single sources there are multiple components fitted in the F606W imaging. This discrepancy demonstrates the need to perform such measurements in bands that trace the underlying old stellar population. We further discuss the impact of this on the {stellar mass - size relation} in Section~\ref{discussion} (see the upper right-hand panel of Figure~\ref{sizemass}). The measured source properties are given in Table~\ref{galprop}.

\begin{table*}
\caption{\label{galprop} Source properties in the cluster SPT-CL J0546$-$5345. }
\begin{centering}
\begin{tabular}{rrrrrrr}
\hline
ID & RA & Dec & M$_{K\rm s}$ & log($r_e$/kpc) & $n$ & log(M$_*$/M$_\odot$)\\
& [deg] & [deg] & [AB mag] & & &\\
(1) & (2) & (3) & (4) & (5) & (6) & (7)\\
\hline
\hline
      15&86.64200&-53.76776&-22.31$\pm$0.74& 0.42$\pm$$^{1.20}_{2.42}$& 3.46$\pm$1.78&10.40$\pm$$^{0.26}_{0.18}$\\
      19&86.62950&-53.76595&-24.39$\pm$0.02& 0.48$\pm$$^{0.10}_{0.10}$& 2.41$\pm$0.05&11.33$\pm$$^{0.11}_{0.14}$\\
      20&86.66529&-53.76681&-22.46$\pm$0.09& 0.41$\pm$$^{0.10}_{0.10}$& 1.45$\pm$0.21&10.58$\pm$$^{0.10}_{0.12}$\\
      21&86.64479&-53.76673&-22.03$\pm$0.16&-0.11$\pm$$^{0.30}_{1.57}$& 4.99$\pm$1.72&10.25$\pm$$^{0.24}_{0.16}$\\
      25&86.64475&-53.76554&-25.33$\pm$0.28& 2.06$\pm$$^{0.10}_{0.10}$&10.00$\pm$1.94&11.80$\pm$$^{0.02}_{0.12}$\\
      26&86.64337&-53.76559&-22.55$\pm$0.14& 0.28$\pm$$^{0.12}_{0.16}$& 2.82$\pm$0.51&10.56$\pm$$^{0.22}_{0.25}$\\
      32&86.65596&-53.76437&-21.63$\pm$0.13& 0.11$\pm$$^{0.10}_{0.11}$& 1.88$\pm$0.50&10.14$\pm$$^{0.13}_{0.22}$\\
      36&86.64292&-53.76368&-22.74$\pm$0.05& 0.13$\pm$$^{0.10}_{0.10}$& 4.60$\pm$0.33&10.74$\pm$$^{0.16}_{0.26}$\\
      37&86.64383&-53.76334&-23.36$\pm$0.15& 0.38$\pm$$^{0.21}_{0.40}$& 4.85$\pm$0.85&11.00$\pm$$^{0.11}_{0.21}$\\
      45&86.63371&-53.76251&-21.78$\pm$0.09&-0.19$\pm$$^{0.10}_{0.10}$& 3.26$\pm$0.87&10.07$\pm$$^{0.04}_{0.04}$\\
      48&86.63496&-53.76229&-22.46$\pm$0.04& 0.05$\pm$$^{0.10}_{0.10}$& 5.14$\pm$0.29&10.59$\pm$$^{0.43}_{0.09}$\\
      50&86.64887&-53.76176&-23.64$\pm$0.08& 0.41$\pm$$^{0.10}_{0.10}$& 7.91$\pm$0.69&11.02$\pm$$^{0.10}_{0.11}$\\
      51&86.63508&-53.76195&-22.75$\pm$0.16& 0.14$\pm$$^{0.10}_{0.10}$&10.00$\pm$5.22&10.63$\pm$$^{0.26}_{0.17}$\\
      53&86.64000&-53.76137&-24.29$\pm$0.14& 0.79$\pm$$^{0.27}_{0.80}$& 4.63$\pm$0.54&11.36$\pm$$^{0.07}_{0.04}$\\
      54&86.65217&-53.76156&-22.45$\pm$0.14& 0.13$\pm$$^{0.70}_{2.13}$& 4.75$\pm$0.76&10.64$\pm$$^{0.23}_{0.24}$\\
      57&86.65371&-53.76140&-24.07$\pm$0.22& 1.44$\pm$$^{0.13}_{0.20}$&10.00$\pm$1.67&11.35$\pm$$^{0.30}_{0.15}$\\
      58&86.65246&-53.76154&-21.31$\pm$0.37& 0.24$\pm$$^{0.10}_{0.10}$& 5.29$\pm$2.72&10.07$\pm$$^{0.33}_{0.30}$\\
      60&86.66146&-53.76012&-24.08$\pm$0.19& 0.77$\pm$$^{0.35}_{2.77}$& 5.76$\pm$0.76&11.29$\pm$$^{0.21}_{0.09}$\\
      61&86.64508&-53.75998&-23.12$\pm$0.11& 0.37$\pm$$^{0.12}_{0.17}$& 2.93$\pm$0.45&10.68$\pm$$^{0.10}_{0.19}$\\
      62&86.65804&-53.75987&-24.20$\pm$0.04& 0.62$\pm$$^{0.10}_{0.10}$& 4.61$\pm$0.17&11.35$\pm$$^{0.08}_{0.03}$\\
      65&86.65371&-53.75984&-22.01$\pm$0.09&-0.04$\pm$$^{0.10}_{0.10}$& 2.71$\pm$0.51&10.35$\pm$$^{0.23}_{0.23}$\\
      66&86.65054&-53.75965&-22.45$\pm$0.26& 0.44$\pm$$^{0.27}_{0.81}$& 7.98$\pm$1.97&10.63$\pm$$^{0.18}_{0.29}$\\
      67&86.65600&-53.75956&-22.09$\pm$0.08& 0.00$\pm$$^{0.10}_{0.10}$& 2.79$\pm$0.44&10.50$\pm$$^{0.18}_{0.25}$\\
      68&86.65879&-53.75959&-21.81$\pm$0.07&-3.55$\pm$$^{2.39}_{0.10}$& 2.98$\pm$1.42&10.27$\pm$$^{0.25}_{0.25}$\\
      69&86.65762&-53.75918&-23.48$\pm$0.03& 0.17$\pm$$^{0.10}_{0.10}$& 4.77$\pm$0.18&10.84$\pm$$^{0.28}_{0.17}$\\
      72&86.65746&-53.75890&-23.38$\pm$0.09& 0.42$\pm$$^{0.10}_{0.10}$& 2.50$\pm$0.24&10.73$\pm$$^{0.36}_{0.11}$\\
      75&86.65708&-53.75909&-22.69$\pm$0.02& 0.30$\pm$$^{0.10}_{0.10}$&10.00$\pm$0.14&10.56$\pm$$^{0.36}_{0.19}$\\
      76&86.65333&-53.75912&-21.13$\pm$0.07&-0.29$\pm$$^{0.10}_{0.10}$& 2.59$\pm$0.71& 9.96$\pm$$^{0.34}_{0.25}$\\
      77&86.65654&-53.75870&-24.90$\pm$0.40& 1.20$\pm$$^{0.10}_{0.10}$& 7.19$\pm$0.46&11.47$\pm$$^{0.05}_{0.03}$\\
      80&86.65696&-53.75884&-21.96$\pm$0.05& 0.27$\pm$$^{0.10}_{0.10}$&10.00$\pm$0.36&10.20$\pm$$^{0.26}_{0.14}$\\
      81&86.63404&-53.75795&-24.03$\pm$0.05& 0.62$\pm$$^{0.10}_{0.10}$& 2.76$\pm$0.13&11.07$\pm$$^{0.04}_{0.06}$\\
      83&86.65275&-53.75854&-22.38$\pm$0.15& 0.20$\pm$$^{0.14}_{0.21}$& 3.95$\pm$0.77&10.48$\pm$$^{0.19}_{0.26}$\\
      85&86.65500&-53.75820&-22.79$\pm$0.04&-0.11$\pm$$^{0.10}_{0.10}$& 2.69$\pm$0.22&10.84$\pm$$^{0.02}_{0.20}$\\
      89&86.65483&-53.75834&-21.27$\pm$0.65&-0.82$\pm$$^{0.10}_{0.10}$& 4.04$\pm$7.42&10.17$\pm$$^{0.24}_{0.31}$\\
      90&86.65742&-53.75806&-21.23$\pm$0.07&-0.19$\pm$$^{0.10}_{0.10}$& 2.51$\pm$0.50&10.23$\pm$$^{0.47}_{0.32}$\\
      91&86.64396&-53.75743&-22.89$\pm$0.04& 0.12$\pm$$^{0.10}_{0.10}$& 2.62$\pm$0.23&10.61$\pm$$^{0.26}_{0.13}$\\
      92&86.65483&-53.75720&-24.25$\pm$0.06& 0.75$\pm$$^{0.10}_{0.10}$& 3.79$\pm$0.19&11.39$\pm$$^{0.11}_{0.11}$\\
      93&86.65162&-53.75729&-23.19$\pm$0.14& 0.53$\pm$$^{0.12}_{0.17}$& 4.57$\pm$0.59&10.97$\pm$$^{0.13}_{0.09}$\\
      97&86.65275&-53.75734&-22.64$\pm$0.09& 0.22$\pm$$^{0.10}_{0.10}$& 4.87$\pm$0.54&10.70$\pm$$^{0.16}_{0.26}$\\
     103&86.63679&-53.75676&-23.74$\pm$0.13& 0.71$\pm$$^{0.11}_{0.15}$& 4.99$\pm$0.55&11.21$\pm$$^{0.05}_{0.20}$\\
     104&86.62417&-53.75684&-22.28$\pm$0.08& 0.27$\pm$$^{0.10}_{0.10}$& 2.21$\pm$0.26&10.29$\pm$$^{0.24}_{0.09}$\\
     107&86.64717&-53.75568&-23.89$\pm$0.02& 0.11$\pm$$^{0.10}_{0.10}$& 2.81$\pm$0.10&11.20$\pm$$^{0.10}_{0.11}$\\
     109&86.64625&-53.75573&-21.58$\pm$0.07&-0.03$\pm$$^{0.10}_{0.10}$& 1.72$\pm$0.31&10.06$\pm$$^{0.18}_{0.17}$\\
     110&86.62558&-53.75568&-23.09$\pm$0.41& 0.85$\pm$$^{0.22}_{0.49}$& 2.91$\pm$0.69&10.64$\pm$$^{0.07}_{0.01}$\\
     111&86.65850&-53.75559&-22.03$\pm$0.31& 0.45$\pm$$^{0.20}_{0.40}$& 1.66$\pm$0.54&10.19$\pm$$^{0.14}_{0.04}$\\
     116&86.65275&-53.75434&-22.97$\pm$0.05& 0.11$\pm$$^{0.10}_{0.10}$& 3.06$\pm$0.25&10.78$\pm$$^{0.24}_{0.17}$\\
     117&86.65058&-53.75426&-22.45$\pm$0.19& 0.29$\pm$$^{0.19}_{0.33}$& 3.68$\pm$0.74&10.63$\pm$$^{0.19}_{0.29}$\\
     121&86.65237&-53.75315&-23.01$\pm$0.07& 0.36$\pm$$^{0.10}_{0.10}$& 3.74$\pm$0.23&10.85$\pm$$^{0.18}_{0.17}$\\
     126&86.64612&-53.75059&-24.70$\pm$0.03& 0.74$\pm$$^{0.10}_{0.10}$& 7.90$\pm$0.16&11.63$\pm$$^{0.04}_{0.08}$\\
\hline
\end{tabular}\\
\end{centering}
Columns: (1) ID; (2) right ascension; (3) declination; (4) $K_s$-band absolute magnitude; (5) {$K_s$-band} effective radius; (6)  {$K_s$-band} S{\'e}rsic index; (7) logarithm of stellar mass. Errors are Monte Carlo simulated: stellar mass within FAST and remaining properties with our own simulations as described in the text.
\end{table*}

\subsection{Error analysis\label{error}}
GALFIT uses a $\chi^2$ minimisation technique to fit profiles and estimate uncertainties under the assumption that an analytical function (in this case, a single S{\'e}rsic profile) is an accurate description of the galaxy being fit, and thus that any residual difference between the model and observed galaxy is solely Poisson noise \citep{Peng2010}. However, this assumption is not valid in many cases, e.g. due to remaining sky structure and contamination by nearby galaxies,  so the residuals are not solely Poisson but also correlated \citep{Haussler2007}. This means that the true uncertainty is underestimated by $\chi^2$ statistics.

In our case we consider the dominant source of error to be structure in the sky background, since GALFIT does a reasonable job of deblending neighbouring galaxies via simultaneous fitting. We consequently estimate uncertainties {for both GeMS and HST imaging} with a Monte Carlo simulation as follows. We insert model galaxies matching the fitted parameters into random locations in a blank patch of sky, and measure the standard error in the recovered parameters. This traces the effect of the large-scale variation in the sky background on the ability of GALFIT to provide a consistent solution. These uncertainties are given in Table~\ref{galprop} and as error bars in Figures~\ref{cmd} and~\ref{sizemass}.

\begin{figure}
\centerline{
\includegraphics[width=1\linewidth]{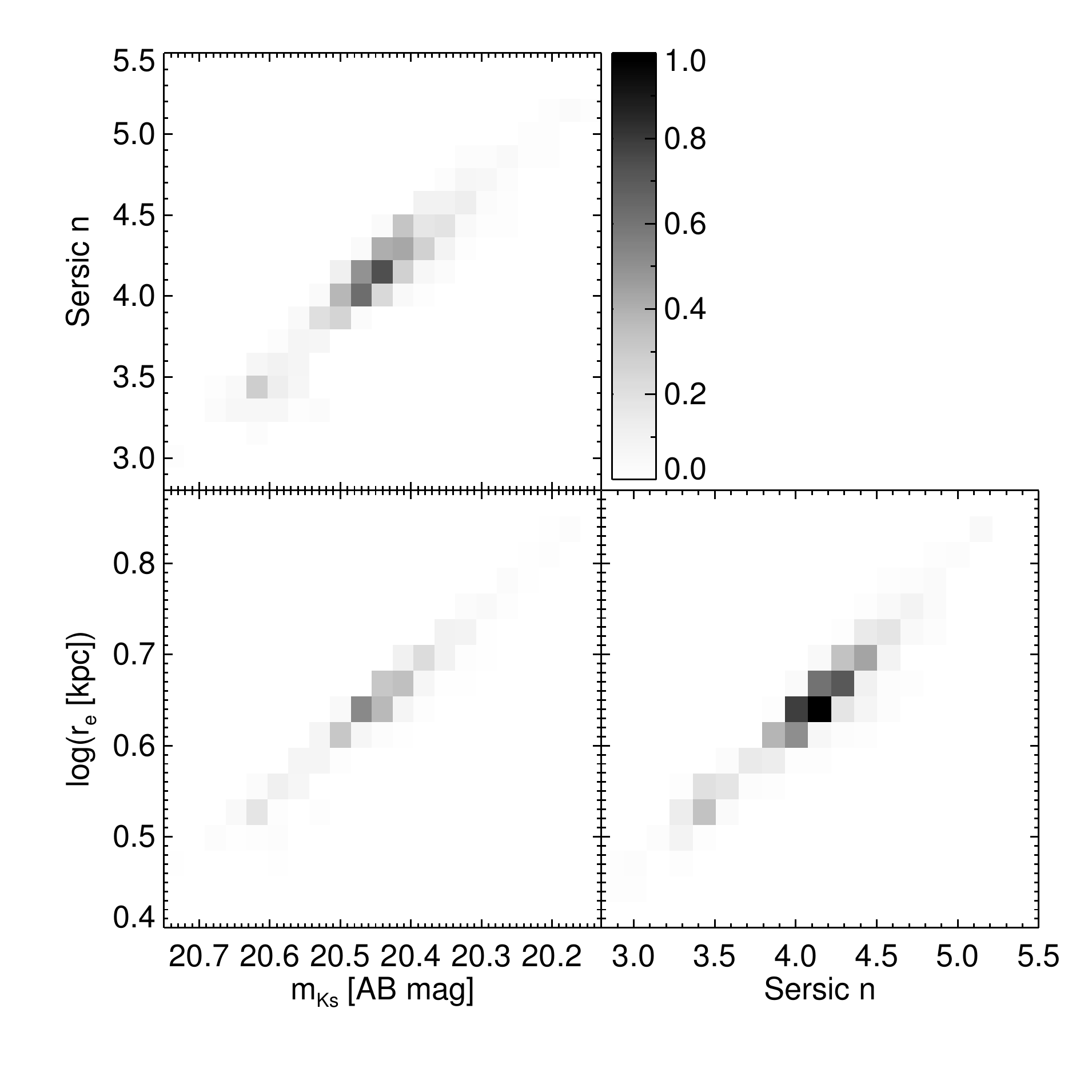}
}
\caption{\label{covariance} Covariance between recovered magnitude, effective radius and S{\'e}rsic index of a Monte Carlo simulated brightest cluster galaxy. 
}
\end{figure}

We show the distribution of recovered parameters of the brightest cluster galaxy in Figure~\ref{covariance}. There is an obvious covariance between magnitude, effective radius and S{\'e}rsic index, such that a smaller magnitude (resulting from a lower sky background and consequently higher flux attributed to the galaxy) corresponds to a higher S{\'e}rsic index and larger effective radius. However, the recovered parameters are strongly clustered around the input model parameters; this is reflected in the low uncertainties in our simulation{: $\sigma _{m_{K_s}} = 0.4$; $\sigma _n = 0.5$; $\sigma _{r_e} = 0.08$.}

\section{The {stellar mass - size relation}}
\label{discussion}

We now investigate the {stellar mass - size relation} measured at high angular resolution in the old stellar population light for SPT-CL J0546$-$5345 cluster members, as presented in Figure~\ref{sizemass} and Table~\ref{slopes}. Our main findings are discussed below.

\subsection{S{\'e}rsic indices} 

\begin{figure}
\centerline{
\includegraphics[width=1\linewidth]{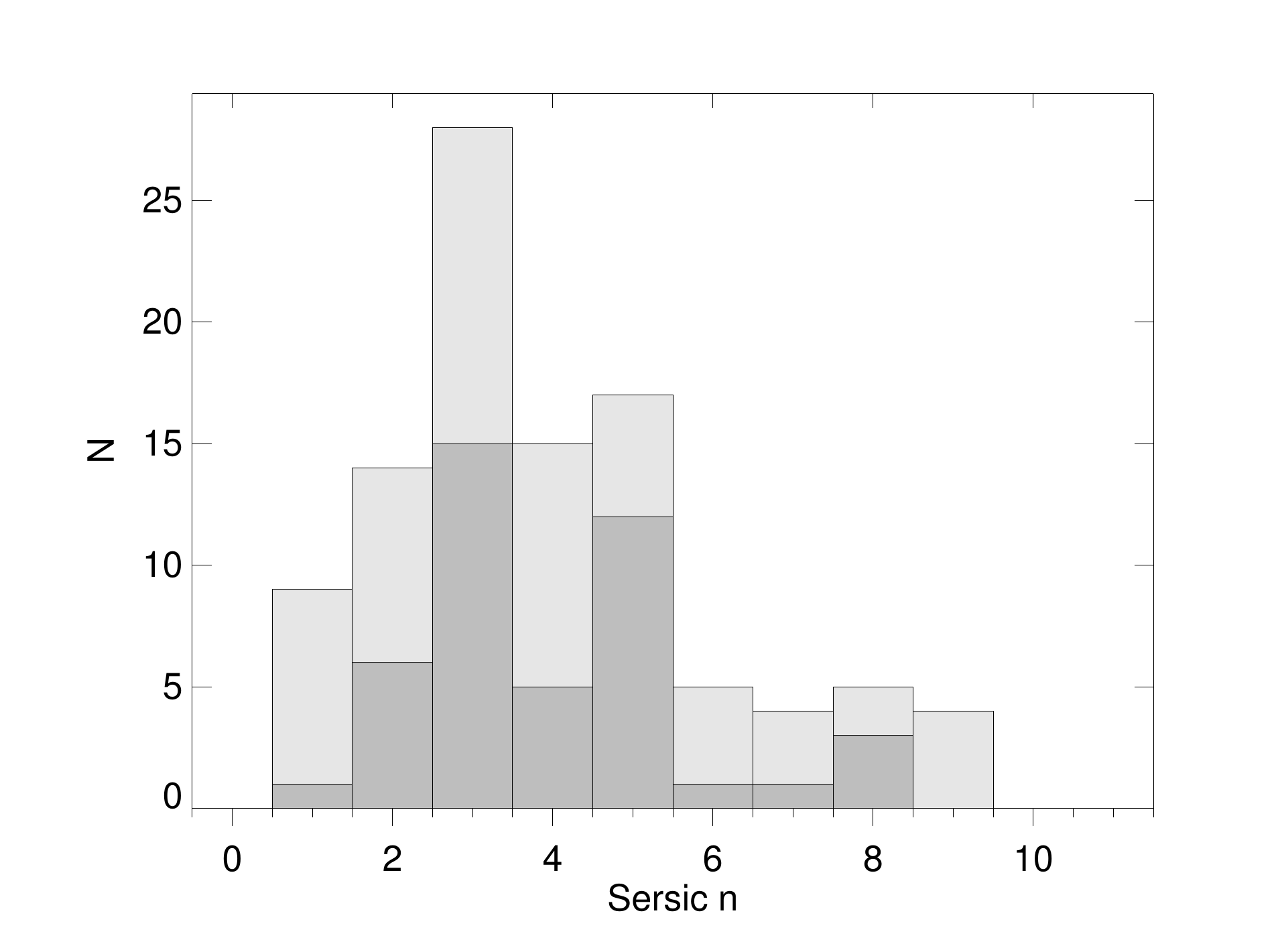}
}
\caption{\label{nhist} Histogram of S{\'e}rsic indices for entire sample (light grey) and cluster members (mid grey). The cluster member sample consists of mostly early-type galaxies, with a median $n$ = 3.8.
}
\end{figure}

In Figure~\ref{nhist} we give the histogram of S{\'e}rsic indices for cluster members and field galaxies. The median for the cluster is $n$ = 3.8 $\pm$ 0.5, while the median for the entire sample is 3.7 $\pm$ 0.5. The large median S{\'e}rsic index for the cluster members indicates early-type galaxies, so in this section we compare with other authors' early-type samples. These are mostly selected to be quiescent, with the exceptions of \citet{Delaye2014} and some datasets in the compilation by \citet{Damjanov2011} which have a morphological selection.

We see no trend of S{\'e}rsic index with cluster-centric radius, indicating no effect of environment on morphology. This is consistent with \citet{HuertasCompany2013} who did not find a significant difference in morphology between field and central galaxies within a population of quiescent early-type galaxies in COSMOS.

\subsection{The {stellar mass - size relation} at $z$~=~1 is offset from that at $z$~=~0.}

\begin{figure*}
\centerline{
\includegraphics[width=1\linewidth]{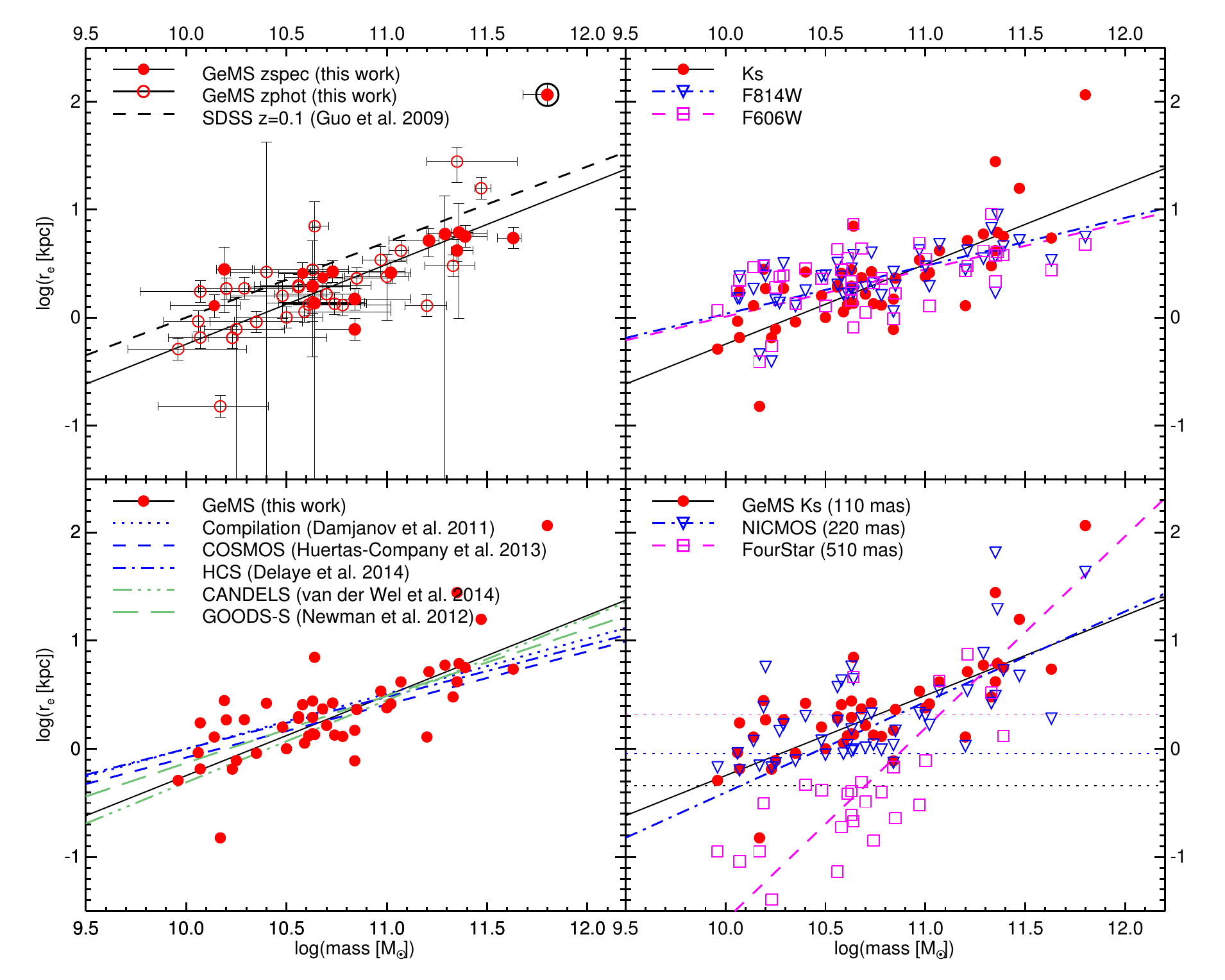}
}
\caption{\label{sizemass} {Stellar mass - size relation} for the cluster SPT-CL J0546$-$5345 compared with other work in the literature. The parameters presented in this Figure are summarised in Table~\ref{slopes}.
{\it Top left:} K$_s$-band {stellar mass - size relation} at $z$~=~1. The relation defined by our cluster members has a slope of $\beta$ = 0.74, consistent with but offset by 0.21 dex from the $z$~=~0 relation shown as the dashed line. Both relations trace the underlying stellar population. The BCG (circled) and the next-largest galaxy appear as outliers above the {stellar mass - size relation}.
{\it Top right:} {Stellar mass - size relation} for the SPT cluster measured in GSAOI K$_s$ (red filled circles and black solid line), HST ACS F814W (rest-frame B-band shown as blue down triangles and dot-dashed line) and F606W (rest-frame U-band as magenta squares and dashed line) bands. Shorter wavelengths give a shallower slope with larger scatter in radius, being affected by UV-bright star-forming knots. 
{\it Bottom left:} Other $z\sim$ 1 samples from the literature, measured in various rest-frame wavelengths. Rest-frame B-band measurements are shown in blue, and rest-frame V-band in green.
{\it Bottom right:} The effect of resolution on the {stellar mass - size relation}. Filled red circles and solid line depict our GeMS K$_s$-band imaging (FWHM $\sim$ 110 mas). Blue down triangles and dot-dashed line show measurements from our imaging smoothed to the resolution of NICMOS (220 mas); magenta squares and dashed line are from smoothing to the resolution of FourStar (510 mas). The horizontal dotted lines indicate the physical size at $z$~=~1 that corresponds to the resolution of each instrument.
}
\end{figure*}

The size-mass distribution for the $z$~=~1 SPT-CL J0546$-$5345 cluster, and best-fitting line of the form log($r_e$)~=~$\kappa + \beta$ log(M$_*$), are indicated in each panel of Figure~\ref{sizemass} as red circles and solid line, and given in Table~\ref{slopes}. In the upper, left-hand plot we show that the $z$~=~1 {stellar mass - size relation} is offset at log(M$_*$) = 11 by 0.21 dex from the present-day SDSS relation \citep[for early-type central galaxies in SDSS,][]{Guo2009}. We calculate the corresponding redshift evolution $\gamma \propto (1+z)^{\alpha}$ of the median mass-normalised size as $\gamma$ = $r_e/(M_*/10^{11}M_\odot)$, and find a slope of $\alpha$ = $-$1.25. We note that our calculation is based on just two data points. The high-redshift data point may be biased since the cluster is massive for its redshift and therefore likely to be more evolved than average. This would potentially bias the measurement towards underestimating the amount of size growth of cluster galaxies between $z$~=~1 and $z$~=~0.
However, our measurement is not discrepant with the literature, which exhibits a wide variation in redshift evolution slope, e.g. $\alpha$ = $-$1.06 in the XDF by \citep{Morishita2014}, $-$1.3 in NEWFIRM \citep{vanDokkum2010}, $-$1.48 in 3D-HST+CANDELS \citep{vanderWel2014}, $-$1.62 in the compilation by \citet{Damjanov2011}. {These works study less evolved structures: the NEWFIRM sample consists of massive groups, while XDF, CANDELS and \citet{Damjanov2011} are field galaxies. Environmental differences notwithstanding, the consistency of our measurement with the literature is not surprising simply because the literature spans such a wide range. The range in } $\alpha$ {is likely driven by} other differences in sample selection (including redshift and galaxy morphology), slopes (by which the samples are mass-normalised) and rest-frame wavelengths (which we discuss below). 

\begin{table*}
\caption{Table of properties of samples presented in Figure~\ref{sizemass}, where log($r_e$) = $\kappa + \beta$ log(M$_*$). 
}
\label{slopes}
\begin{tabular}{llllll}
\hline
Sample & z & rest-frame & $\beta$ & $\kappa$ & Reference\\
\hline 
\hline
GeMS Ks &1.067&Y&0.74$\pm$0.06&-7.65$\pm$0.64&This work\\
SDSS&0.1&$z$&0.70$\pm$0.05&-7.00$\pm$0.30*&Guo et al. 2009\\
\hline
F814W&1.067&B&0.45$\pm$0.04&-4.42$\pm$0.44&This work\\
F606W&1.067&U&0.44$\pm$0.08&-4.38$\pm$0.50&This work\\
\hline
Compilation&0.8 $<z<$ 1.4&B&0.51$\pm$0.06&-5.10$\pm$0.10*&Damjanov et al. 2011\\
COSMOS&0.8 $<z<$ 1.0&B&0.49$\pm$0.04&-4.98$\pm$0.41
&Huertas-Company et al. 2013\\
HCS&0.9 $<z<$ 1.1&B&0.48$\pm$0.08&-4.80$\pm$0.30&Delaye et al. 2014\\
3D-HST+CANDELS&1.25&V&0.76$\pm$0.04&-7.91$\pm$0.01& van der Wel et al. 2014\\
GOODS-S&1.0 $<z<$ 1.5&V&0.62$\pm$0.09&-6.33$\pm$0.02& Newman et al. 2012\\
\hline
NICMOS (220 mas)&1.067&Y&0.84$\pm$0.06&-8.76$\pm$0.60&This work\\
FourStar (510 mas)&1.067&Y&1.77$\pm$0.21&-19.32$\pm$2.25&This work\\
\hline 
\hline
\end{tabular}\\
Notes: (*) intercept not given in reference; estimated by eye.
\end{table*}

\subsection{Effect of rest-frame wavelength}

For observations taken in the rest-frame UV, one would expect that measured $r_e$ is affected by the clumpy star-forming regions that dominate the flux at those wavelengths. This should be particularly the case for rest-frame $U$-band, which falls almost entirely blueward of the 4000\AA\ break, and $B$-band, which straddles it. 

At the cluster redshift of $z$~=~1.067, our HST/ACS F606W, HST/ACS F814W and $K_{\rm s}$ imaging correspond to rest-frame observations in approximately the $U$, $B$ and $Y$ bands (2930\AA, 3940\AA\ and 1.05$\,\mu$m). We investigate the effect of rest-frame wavelength on the slope of the {stellar mass - size relation} by comparing the cluster member galaxies' effective radii in these three bands. The resulting {stellar mass - size relation}s are illustrated in the upper, right-hand panel of Figure~\ref{sizemass}, and exhibit slopes of $\beta$ = 0.44 $\pm$ 0.08, 0.45 $\pm$ 0.04 and 0.74 $\pm$ 0.06 respectively (Table~\ref{slopes}). The rest-frame $U$ and $B$ imaging yield significantly shallower slopes than rest-frame $Y$, indicating that the bluer images are indeed affected by clumpy star-formation events, particularly in lower mass galaxies. However, there is no significant wavelength effect on the zero-point of the measured {stellar mass - size relation}s, with an offset in log(r$_e$) at log(M$_*$) = 11 of just $-$0.04 dex between $Y$ and $B$ and +0.03 dex between $Y$ and $U$ bands. The rest-frame $B$ and $U$ relations have similar scatter to rest-frame $Y$.
The fact that this analysis is based on a consistent sample rules out any sample differences (such as S{\'e}rsic index, mass, environment, redshift, star formation rate) as being responsible for the difference in slope. The mechanism for the shallower slopes from the two bluer images is that in the case of some galaxies, star-forming clumps at the edges of galaxies serve to overestimate their measured $r_e$ (preferentially for the smallest galaxies); while in other cases, multiple star-forming clumps within a galaxy are measured as multiple galaxies with consequently underestimated $r_e$ (preferentially for the largest galaxies).\footnote{{While many of the cluster members in this work are located on the red sequence, a significant proportion have bluer colours with clumps or a frosting of star formation. It is these galaxies in particular whose sizes can be under- or overestimated in this way.}}
 
We further illustrate the effect of rest-frame wavelength by comparing with other $z$~=~1 samples from the literature in the lower, left-hand panel of Figure~\ref{sizemass}, where rest-frame $B$-band measurements are shown in blue and rest-frame $V$-band in green; these are also given in Table~\ref{slopes}. The $B$-band {stellar mass - size relation}s measured in a compilation by \citet{Damjanov2011}, in COSMOS by \citet{HuertasCompany2013} and in HCS by \citet{Delaye2014} ($\beta$ = 0.51 $\pm$ 0.06, 0.49 $\pm$ 0.04, 0.48 $\pm$ 0.08 respectively) are all consistent with our rest-frame $U$-band and $B$-band slopes. On the other hand, the rest-frame $V$-band {stellar mass - size relation}s measured in 3D-HST+CANDELS by \citet{vanderWel2014} and GOODS-S by \citet{Newman2012} ($\beta$ = 0.76 $\pm$ 0.04 and 0.62 $\pm$ 0.09 respectively) more closely match our rest-frame $Y$-band $\beta$. From this agreement we draw the conclusion that observations made in the rest-frame $V$-band, which falls entirely redward of the 4000\AA\ break, are not greatly affected by star-forming clumps. 

\subsection{Effect of resolution}

When measuring the sizes of galaxies, one requires sufficient resolution that the sources are resolved in the imaging. Given that high-redshift galaxies are very compact, one would expect that resolution is particularly important when measuring the {stellar mass - size relation} at high redshift.

We investigate the impact of substituting our $\sim$110 mas resolution $K_s$-band GeMS imaging with the effective resolution of imaging from alternative instruments at similar wavelengths: HST/NICMOS\footnote{We do not compare to WFC3 as that instrument does not have a $\sim K$-band filter.}, with diffraction-limited FWHM = 220 mas in the F222M filter ($\sim K$-band) \citep{Scoville2000}, and Magellan/FourStar, with seeing-limited FWHM = 510 mas in $K_s$-band \citep{Lee2012}. We smooth our $K_s$-band image with a Gaussian filter to simulate the resolution of two such data-sets. The resulting {stellar mass - size relation}s are given in the lower, right-hand panel of Figure~\ref{sizemass}, and in Table~\ref{slopes}. The slope of the NICMOS-resolution {stellar mass - size relation} is not significantly different from GeMS at $\beta$ = 0.84 $\pm$ 0.06, and the zero-point is just 0.01 dex lower in log(r$_e$) at log(M$_*$) = 11.
This broad agreement suggests that the resolution delivered by current space-based instruments may be sufficient for accurate measurement of the high-redshift {stellar mass - size relation} in the rest-frame light of the old stellar population, even though the sizes of the smallest galaxies are below the resolution limit. We point out that the narrow field of view of NICMOS (19\arcsec.5 x 19\arcsec.5) makes this measurement observationally expensive for large sample sizes. Similarly, there is agreement between the slopes measured at 180 mas in the observed $H$-band of 3D-HST+CANDELS and GOODS-S (green best fitting lines in lower, left-hand panel of Figure~\ref{sizemass}) and that of our higher resolution GeMS relation. However, while this implies that HST/ACS $H$-band observations are adequate for this type of measurement at $z$~=~1, they would not suffice for significantly higher redshifts, at which this filter straddles the 4000\AA\ break. 

In contrast to the space-based resolutions presented above, the {stellar mass - size relation} measured at seeing-limited ground-based resolution (FourStar) is significantly steeper than our GeMS {stellar mass - size relation}, indicating that the relation cannot be accurately measured from the ground without the use of adaptive optics. This is as expected, since the true $r_e$ falls well below the FourStar resolution for all but the largest galaxies in our sample.

\subsection{The slope of the {stellar mass - size relation} at $z$~=~1 is consistent with that at $z$~=~0.}

Major mergers have been proposed as a mechanism for the observed evolution in the {stellar mass - size relation}. These are known to cause massive galaxies to grow more quickly than less massive ones \citep{Khochfar2006}. Thus if major mergers were a dominant source of the evolution in the {stellar mass - size relation}, then the most massive galaxies would experience the most rapid growth, so that the slope of the {stellar mass - size relation} would be steeper at the present day than at $z$~=~1. However, our $z$~=~1 slope of $\beta$ = 0.74 $\pm$ 0.06 is consistent with the $z$~=~0.1 slope of $\beta$ = 0.70 measured by \cite{Guo2009} (upper, left-hand panel of Figure~\ref{sizemass}). The two samples have very similar wavelength ranges (rest-frame $Y$-band and $z$-band), which both trace the old stellar population; thus this comparison is not affected by clumpy star formation, but instead demonstrates that the slope of the {stellar mass - size relation} is constant from $z$~=~1 to today. Consequently, massive galaxies typically grow at the same rate as smaller galaxies, so that major mergers are not responsible for the evolution in the {stellar mass - size relation} between $z$~=~1 and today\footnote{Interestingly, our BCG (and the second-largest galaxy) has a somewhat enhanced size above the {stellar mass - size relation}, suggesting that it may have had an increased history of major mergers by virtue of its location at the bottom of the potential well. Also note that our analysis applies to the most massive evolved clusters at $z$~=~1, and not to galaxies in more typical clusters at that redshift, in which there is strong evidence for major mergers driving galaxy assembly. The timing of the growth may also be different in the field, since the cluster environment accelerates galaxy evolution.}. 

Our conclusion corresponds with other authors' findings spanning this redshift range. The general understanding built by the literature is that the majority of structures were in place by $z$~=~1 \citep{Papovich2005}, and that major mergers at all galaxy masses are rare since then \citep[see also][]{Bundy2009,Lotz2011}. At lower redshifts this is evidenced by a constant size-mass slope at lower redshifts, e.g. \citet{Delaye2014} found the slope to be constant up to $z\sim$ 1.4. For galaxies in COSMOS at 0.2 $<z<$ 1.1, \citet{HuertasCompany2013} measured a constant slope for a sample of passive galaxies but an evolving slope for galaxies with $n>$ 2.5, indicating that the sample selection affects the measured size growth. At higher redshifts an evolving size-mass slope is the result of high-mass galaxies experiencing more rapid growth in $r_e$ than low-mass ones, due to a greater number of major mergers for high-mass galaxies, e.g. \citet{Ryan2012}. \citet{Brodwin2013} also found significant major merger rates in clusters above $z\sim$ 1.3, as did \citet{Mancone2010} in clusters at $z>$ 1.4. A few authors measure a consistent size-mass slope into high redshifts; \citet{Newman2012} and \citet{vanderWel2014} showed no evolution in slope with redshift between 0.4 $<z<$ 2.5 and 0.25 $<z<$ 2.75 respectively, for measurements made in the rest-frame $V$ band in both cases. \citet{Damjanov2011} also observed a consistent slope over 0.2 $<z<$ 2.7. However, the rest-frame wavelength for that sample decreases with increasing redshift, and consequently the higher-redshift bins may be subject to the same biases demonstrated in our rest-frame $U$-band imaging{, though if we were to (incorrectly) compare our rest-frame $U$-band relation with the rest-frame $z$-band one at $z$ = 0 we would in fact detect an evolving slope and attribute the growth to major mergers. The consistent slope above $z >$ 1 observed by \citet{Damjanov2011}, \citet{Newman2012} and \citet{vanderWel2014}} is at odds with the prevalence of major mergers at early times as understood from theoretical works such as \citet{Khochfar2001,Fakhouri2010,Rodriguez-Gomez2015}. Our observation that the size-mass slope is constant since $z\sim$ 1 indicates that the observed size evolution since that time does not depend on stellar mass; it is thus consistent with the framework where major mergers are not responsible for the observed growth at late times.

The two remaining proposed channels for this growth are minor mergers and adiabatic expansion. First we consider the case where a series of minor mergers is responsible for the size growth. It is common to parametrize the growth efficiency as $\alpha = d {\rm log} r_e / d {\rm log } M_*$. The dependence of growth efficiency on mass ratio $\mu$ is given by 
\begin{equation}
\alpha = 2 - \frac{{\rm log}(1 + \mu^{2-\beta})}{{\rm log}(1 + \mu)},
\end{equation} where $\beta$ is the slope of the {stellar mass - size relation} \citep{Newman2012}. For major mergers of equal-sized galaxies, $\mu$ = 1 so that the minimum $\alpha$ = 1. For minor mergers the literature contains a wide range in growth efficiency, for example $\alpha$ = 1.3 \citep{Nipoti2009}, $\alpha$~=~1.6 \citep{Newman2012} and $\alpha \gtrsim$ 2.7 \citet{Bezanson2009}. Our measured $\beta$~=~0.74~$\pm$~0.06 yields a maximum $\alpha$~=~1.43~$\pm$~0.08 assuming the same minor merger mass ratio $\mu$ =0.1. This is a somewhat lower growth efficiency than \citet{Newman2012} due to the steeper slope we measure, but more efficient than the \citet{Nipoti2009} result for minor mergers. It may be sufficiently efficient to explain the amount of observed growth from $z$~=~1 to today (though not from $z$~=~2 to 1) \citep{Newman2012}. With regards to the requirement to match the observed constancy of slope, we note that the mass ratio $\mu$ was shown to be independent of stellar mass of the larger galaxy in the low-redshift GAMA survey \citep{Robotham2014}, so the growth efficiency is also independent of stellar mass. Size growth by minor mergers therefore appears to be consistent with the slope of the size-mass remaining unchanged with redshift.

Next we consider the case where adiabatic expansion due to rapid mass loss is responsible for the size growth. In this case, expansion scales in proportion to mass lost \citep{RangoneFigueroa2011}, so for a constant size-mass slope each galaxy must experience the same amount of mass loss per unit host galaxy mass. Mass loss driven by AGN winds is consistent with this case, as it is known that outflow energy scales with luminosity of the black hole \citep{Birzan2004,King2013,Heckman2014}, and therefore with the host galaxy mass \citep{Ferrarese2000,Gebhardt2000}.
However, it does not hold for mass loss driven by supernovae, since the specific supernova rate is not constant with stellar mass \citep[smaller galaxies have higher supernova rates,][]{Li2011}. Further to this, \citet{Hopkins2010b} note that adiabatic expansion caused by stellar winds is only sufficient to cause a size increase of 20\%, much less than the observed size increase. \citet{Damjanov2009} also concluded that adiabatic expansion due to stellar winds was unlikely. This suggests that adiabatic expansion due to supernova-driven mass loss cannot be responsible for the redshift evolution of the {stellar mass - size relation}, though adiabatic expansion due to AGN-driven mass loss may.

\section{Summary and Conclusions}
We present the first high angular-resolution measurement of the rest-frame $Y$-band {stellar mass - size relation} for the galaxy cluster SPT-CL J0546$-$5345 at $z$~=~1.067.
We demonstrate our strategies for addressing the data processing challenges associated with these complex imaging data in order to achieve a final stacked image with a mean PSF FWHM~$\sim$~110~mas (corresponding to a radius of 450 pc at the cluster redshift), with a uniform sky subtraction free from strong residual images that echo faint field sources.
The spatial variation of the PSF is modelled as a two-dimensional Moffat profile fit to known stars across the field and interpolated at the location of each measured galaxy.

Forty-nine cluster member galaxies are detected, with median S{\'e}rsic index $n$ = 3.8 $\pm$ 0.5. The {stellar mass - size relation} at $z\sim$ 1 is offset from that at $z\sim$ 0 by 0.21 dex, corresponding to a size evolution of 
$\gamma \propto (1+z)^{-1.25}$, which is consistent with previous results for minor mergers \citep{Nipoti2009}.
 
The {stellar mass - size relation} exhibits a slope of $\beta$ = 0.74 $\pm$ 0.06, consistent with the local slope of $\beta$ = 0.70 $\pm$ 0.05 reported for quiescent low redshift cluster galaxies \citep{Guo2009}. This suggests that the cluster SPT-CL J0546$-$5345, an extremely massive cluster at $z\sim$ 1, has ceased its early, rapid growth dominated by major mergers, leaving the galaxies to increase in size via minor mergers and/or adiabatic expansion due to AGN mass loss winds.

The {stellar mass - size relation} for the cluster members is also measured in the rest-frame $B$ and $U$ bands from archival $HST$/ACS F814W and F606W imaging. We show that those measurements are contaminated by knots of star formation that affect the light profiles. Galaxy effective radii are preferentially overestimated for low-mass systems and preferentially underestimated for high-mass systems when measured in the rest-frame UV wavelengths that are blueward of or straddle the $\lambda$ = 4000\AA\ break, with a stronger bias at shorter wavelengths. 
The effect of this is that the slope of the {stellar mass - size relation} is severely affected by the rest-frame wavelength range at which it is measured, although the zero-point is not affected such that a 10$^{11}$ M$_\odot$ galaxy has an effective radius of 2.45 kpc, consistent with the literature. This is a vivid illustration of the necessity of performing size measurements in the rest-frame underlying old stellar population in order to avoid bias by clumpy regions of star formation.

We measure the {stellar mass - size relation} after Gaussian smoothing to typical imaging resolutions obtained with diffraction-limited HST/NICMOS $K$-band and seeing-limited Magellan/FourStar $K_s$-band. The NICMOS resolution gives a result consistent with our GSAOI imaging, but the seeing-limited FourStar resolution gives a significantly steeper slope. Hence the {stellar mass - size relation} cannot be accurately measured at $z \gtrsim$ 1 from the ground without the use of AO.
	
Having demonstrated the potential of wide field AO observations to characterise the underlying stellar populations in high redshift cluster galaxies, in future work we will present additional clusters across the redshift range 1 $<z<$ 2 to assess the mechanisms for galaxy growth across this key epoch.

\section*{Acknowledgements}

Based in part on observations obtained at the Gemini Observatory, which is operated by the 
Association of Universities for Research in Astronomy, Inc., under a cooperative agreement 
with the NSF on behalf of the Gemini partnership: the National Science Foundation 
(United States), the National Research Council (Canada), CONICYT (Chile), the Australian 
Research Council (Australia), Minist\'{e}rio da Ci\^{e}ncia, Tecnologia e Inova\c{l}\~{a}o 
(Brazil) and Ministerio de Ciencia, Tecnolog\'{i}a e Innovaci\'{o}n Productiva (Argentina).

Elements of the data presented in this paper were obtained from the Mikulski Archive for Space Telescopes (MAST). STScI is operated by the Association of Universities for Research in Astronomy, Inc., under NASA contract NAS5-26555. Support for MAST for non-HST data is provided by the NASA Office of Space Science via grant NNX09AF08G and by other grants and contracts.

This publication makes use of data products from the Two Micron All Sky Survey, which is a joint project of the University of Massachusetts and the Infrared Processing and Analysis Center/California Institute of Technology, funded by the National Aeronautics and Space Administration and the National Science Foundation.

This research has made use of NASA's Astrophysics Data System.

This research has made use of the VizieR catalogue access tool, CDS, Strasbourg, France. The original description of the VizieR service was published in A\&AS 143, 23.

This research was conducted with the support of Australian Research Council DP130101667.
We thank the International Telescope Support Office and Gemini Observatory for contributing travel funding.

\clearpage
\onecolumn
\begin{centering}
\begin{longtable}{lllllllllllll}
\caption{Source identifications in the field of SPT-CL J0546$-$5345.
\label{cat}}\\
\hline
\multicolumn{1}{l}{ID}&\multicolumn{1}{l}{RA}&\multicolumn{1}{l}{Dec}&\multicolumn{1}{l}{$z_{\rm GMOS}$}&\multicolumn{1}{l}{$z_{\rm B10}$}&\multicolumn{1}{l}{$z_{\rm R14}$}&\multicolumn{1}{l}{spec}&\multicolumn{1}{l}{phot}&\multicolumn{1}{l}{m$_{K_{\mathrm s}}$}&\multicolumn{1}{l}{m$_{\mathrm F814W}$}&\multicolumn{1}{l}{m$_{\mathrm F606W}$}&\multicolumn{1}{l}{(b/a)}&\multicolumn{1}{l}{(PA)}\\
\multicolumn{1}{l}{}&\multicolumn{1}{l}{[deg]}&\multicolumn{1}{l}{[deg]}&\multicolumn{1}{l}{}&\multicolumn{1}{l}{}&\multicolumn{1}{l}{}&\multicolumn{1}{l}{}&\multicolumn{1}{l}{}&\multicolumn{1}{l}{[AB mag]}&\multicolumn{1}{l}{[AB mag]}&\multicolumn{1}{l}{[AB mag]}&\multicolumn{1}{l}{}&\multicolumn{1}{l}{[deg]}\\
\multicolumn{1}{l}{(1)}&\multicolumn{1}{l}{(2)}&\multicolumn{1}{l}{(3)}&\multicolumn{1}{l}{(4)}&\multicolumn{1}{l}{(5)}&\multicolumn{1}{l}{(6)}&\multicolumn{1}{l}{(7)}&\multicolumn{1}{l}{(8)}&\multicolumn{1}{l}{(9)}&\multicolumn{1}{l}{(10)}&\multicolumn{1}{l}{(11)}&\multicolumn{1}{l}{(12)}&\multicolumn{1}{l}{(13)}\\
\hline
\hline
\endfirsthead
\caption{Source identifications in the field of SPT-CL J0546$-$5345.}\\
\hline
\multicolumn{1}{l}{ID}&\multicolumn{1}{l}{RA}&\multicolumn{1}{l}{Dec}&\multicolumn{1}{l}{$z_{\rm GMOS}$}&\multicolumn{1}{l}{$z_{\rm B10}$}&\multicolumn{1}{l}{$z_{\rm R14}$}&\multicolumn{1}{l}{spec}&\multicolumn{1}{l}{phot}&\multicolumn{1}{l}{m$_{K_{\mathrm s}}$}&\multicolumn{1}{l}{m$_{\mathrm F814W}$}&\multicolumn{1}{l}{m$_{\mathrm F606W}$}&\multicolumn{1}{l}{(b/a)}&\multicolumn{1}{l}{(PA)}\\
\multicolumn{1}{l}{}&\multicolumn{1}{l}{[deg]}&\multicolumn{1}{l}{[deg]}&\multicolumn{1}{l}{}&\multicolumn{1}{l}{}&\multicolumn{1}{l}{}&\multicolumn{1}{l}{}&\multicolumn{1}{l}{}&\multicolumn{1}{l}{[AB mag]}&\multicolumn{1}{l}{[AB mag]}&\multicolumn{1}{l}{[AB mag]}&\multicolumn{1}{l}{}&\multicolumn{1}{l}{[deg]}\\
\multicolumn{1}{l}{(1)}&\multicolumn{1}{l}{(2)}&\multicolumn{1}{l}{(3)}&\multicolumn{1}{l}{(4)}&\multicolumn{1}{l}{(5)}&\multicolumn{1}{l}{(6)}&\multicolumn{1}{l}{(7)}&\multicolumn{1}{l}{(8)}&\multicolumn{1}{l}{(9)}&\multicolumn{1}{l}{(10)}&\multicolumn{1}{l}{(11)}&\multicolumn{1}{l}{(12)}&\multicolumn{1}{l}{(13)}\\
\hline
\hline
\endhead
\hline
\multicolumn{13}{r}{Continued on next page}
\endfoot
\endlastfoot
       3&86.63450&-53.77259&   -&   -&   -&       0&       0&22.03&23.32&23.51&0.74& -4.22\\
       4&86.63946&-53.77115&   -&   -&   -&       0&       0&22.05&25.28&25.28&0.67& 71.80\\
       5&86.62975&-53.77023&   -&   -&   -&       0&       0&22.98&23.70&23.51&0.77& 88.35\\
       6&86.64387&-53.76906&   -&   -&   -&       0&       0&20.30&23.92&24.48&0.80&-29.73\\
       7&86.63387&-53.76948&   -&   -&   -&       0&       0&23.12&25.99&25.17&0.99&  9.32\\
       8&86.65075&-53.76931&   -&   -&   -&       0&       0&22.64&25.25&24.97&0.54& 52.16\\
       9&86.66058&-53.76859&0.0000&   -&   -&       0&       0&18.67&19.66&21.46&0.04&-54.42\\
      10&86.65396&-53.76834&   -&   -&   -&       0&       0&21.40&25.29&25.53&0.74& 64.71\\
      11&86.64921&-53.76823&   -&   -&   -&       0&       0&21.62&23.07&23.53&0.57&-71.50\\
      12&86.66296&-53.76823&   -&   -&   -&       0&       0&22.23&  -&24.48&0.85& 19.31\\
      13&86.65008&-53.76790&   -&   -&   -&       0&       0&21.69&24.65&25.46&0.33&-47.35\\
      14&86.65562&-53.76770&   -&   -&   -&       0&       0&21.57&23.46&23.93&0.91&-44.81\\
      15&86.64200&-53.76776&   -&   -&   -&       0&       0&21.98&24.31&24.97&0.83&-59.64\\
      16&86.65179&-53.76768&   -&   -&   -&       0&       0&21.22&  -&25.80&0.23&-83.54\\
      17&86.64246&-53.76756&0.5083&   -&   -&       0&       0&22.10&23.09&23.73&0.37&-75.11\\
      18&86.63133&-53.76740&   -&   -&   -&       0&       1&23.14&24.72&25.93&0.50&-69.23\\
      19&86.62950&-53.76595&   -&   -&   -&       0&       0&19.90&22.59&23.54&0.24& 12.39\\
      20&86.66529&-53.76681&   -&   -&   -&       1&       0&21.84&23.67&24.04&0.92& 37.34\\
      21&86.64479&-53.76673&   -&   -&   -&       0&       0&22.27&24.52&24.88&0.77& 68.22\\
      22&86.66071&-53.76615&0.0000&   -&   -&       0&       0&20.82&21.80&24.31&0.45& 75.93\\
      23&86.65296&-53.76576&   -&   -&   -&       0&       0&21.46&24.51&25.63&0.75& -1.28\\
      24&86.63625&-53.76579&   -&   -&   -&       0&       0&22.18&24.73&25.91&0.34& 30.69\\
      25&86.64475&-53.76554&   -&1.0567&1.0567&       1&       1&18.97&22.62&23.96&0.66& 61.40\\
      26&86.64337&-53.76559&   -&   -&   -&       0&       1&21.75&24.13&24.76&1.00&-47.93\\
      27&86.65246&-53.76490&   -&   -&   -&       0&       0&21.04&22.66&24.47&0.76& 54.94\\
      28&86.63808&-53.76498&   -&   -&   -&       0&       0&21.45&25.12&27.25&0.42&-19.17\\
      29&86.64354&-53.76468&   -&   -&   -&       0&       0&20.70&24.44&26.17&0.44& 75.87\\
      30&86.63475&-53.76409&   -&   -&   -&       0&       0&20.53&21.09&22.82&0.66& 53.31\\
      31&86.65746&-53.76412&   -&   -&   -&       0&       1&20.43&23.32&25.20&0.88&-71.70\\
      32&86.65596&-53.76437&   -&   -&   -&       1&       0&22.66&24.21&25.61&0.80&-87.77\\
      33&86.65979&-53.76429&   -&   -&   -&       0&       0&22.28&25.36&25.37&0.66& -2.76\\
      34&86.64917&-53.76434&   -&   -&   -&       0&       0&23.62&26.05&25.65&1.00& 10.43\\
      35&86.66371&-53.76415&   -&   -&   -&       0&       0&22.97&40.02&26.84&0.25& 29.70\\
      36&86.64292&-53.76368&   -&   -&   -&       0&       1&21.56&23.93&25.30&0.53&-29.25\\
      37&86.64383&-53.76334&   -&   -&   -&       0&       1&20.93&23.25&24.60&0.91&-49.55\\
      38&86.63292&-53.76362&   -&   -&   -&       0&       0&23.57&  -&26.22&0.74& 44.60\\
      39&86.65192&-53.76351&   -&   -&   -&       0&       0&21.84&24.75&27.09&0.68&-24.71\\
      41&86.64442&-53.76290&   -&   -&   -&       0&       0&20.75&23.60&24.15&0.77& 72.54\\
      42&86.65925&-53.76309&   -&   -&   -&       0&       0&22.54&25.30&26.43&0.99&  3.08\\
      43&86.64862&-53.76270&   -&   -&   -&       0&       0&20.23&23.65&24.10&0.86&-56.75\\
      44&86.65725&-53.76229&   -&   -&   -&       0&       0&21.28&22.79&24.28&0.45&-48.11\\
      45&86.63371&-53.76251&   -&   -&   -&       0&       1&22.52&24.55&23.51&0.54& 44.36\\
      46&86.63575&-53.76256&   -&   -&   -&       0&       1&23.96&25.23&25.45&0.60&-83.31\\
      48&86.63496&-53.76229&   -&   -&   -&       0&       1&21.84&24.11&23.74&0.37& 75.83\\
      49&86.63675&-53.76223&   -&   -&   -&       0&       0&21.92&  -&23.41&0.53&-58.44\\
      50&86.64887&-53.76176&   -&1.0710&1.0042&       1&       1&20.66&23.09&24.13&0.60&  5.90\\
      51&86.63508&-53.76195&   -&   -&   -&       0&       0&21.55&24.15&24.93&0.56& 38.91\\
      52&86.63887&-53.76151&   -&   -&   -&       0&       0&20.77&23.00&23.93&0.51& 58.25\\
      53&86.64000&-53.76137&   -&1.0775&1.0775&       1&       1&20.01&22.23&23.53&0.71&-46.01\\
      54&86.65217&-53.76156&   -&   -&   -&       1&       1&21.85&24.54&25.85&0.75& 85.07\\
      55&86.64687&-53.76159&   -&   -&   -&       0&       0&21.17&  -&  -&0.43& 88.34\\
      56&86.63533&-53.76159&   -&   -&   -&       0&       0&22.71&25.10&24.97&0.79&-38.11\\
      57&86.65371&-53.76140&   -&   -&   -&       0&       1&20.23&24.12&24.71&0.53&-65.90\\
      58&86.65246&-53.76154&   -&   -&   -&       0&       0&22.99&25.24&26.25&0.91& 19.41\\
      59&86.62962&-53.76109&   -&   -&   -&       0&       0&21.57&24.18&25.21&0.74&-50.41\\
      60&86.66146&-53.76012&1.0693&   -&   -&       1&       1&20.22&23.11&24.70&0.88&-72.32\\
      61&86.64508&-53.75998&0.9331&   -&   -&       1&       0&21.18&  -&23.62&0.65& 11.10\\
      62&86.65804&-53.75987&   -&   -&   -&       1&       1&20.10&23.02&24.56&0.92&-87.30\\
      63&86.66200&-53.76004&   -&   -&   -&       0&       1&23.46&24.71&25.60&0.90&-81.43\\
      64&86.63596&-53.75962&   -&   -&   -&       0&       0&20.68&23.62&23.98&0.55&-34.74\\
      65&86.65371&-53.75984&   -&   -&   -&       0&       1&22.29&24.37&25.63&0.99&-31.52\\
      66&86.65054&-53.75965&   -&   -&   -&       0&       0&21.84&24.23&25.43&0.70& 19.55\\
      67&86.65600&-53.75956&   -&   -&   -&       0&       1&22.21&23.97&26.41&0.94& 24.82\\
      68&86.65879&-53.75959&   -&   -&   -&       0&       0&22.48&24.65&25.71&0.04& 22.02\\
      69&86.65762&-53.75918&0.9954&   -&   -&       1&       1&20.81&23.15&  -&0.46& 40.95\\
      70&86.63875&-53.75868&0.0000&   -&   -&       0&       0&18.06&20.20&22.21&0.63&-13.64\\
      71&86.65158&-53.75943&   -&   -&   -&       0&       0&22.11&24.96&26.72&0.78&-17.88\\
      72&86.65746&-53.75890&0.9954&   -&   -&       1&       1&20.92&23.32&  -&0.82& 39.76\\
      73&86.63637&-53.75920&   -&   -&   -&       0&       0&22.95&25.39&25.94&0.90&-73.38\\
      74&86.64667&-53.75915&   -&   -&   -&       0&       0&22.18&  -&24.73&0.35& 75.89\\
      75&86.65708&-53.75909&   -&   -&   -&       0&       1&21.61&24.05&  -&0.50&-52.48\\
      76&86.65333&-53.75912&   -&   -&   -&       0&       0&23.16&25.23&26.21&0.96& -1.82\\
      77&86.65654&-53.75870&   -&   -&   -&       0&       1&19.40&22.79&  -&0.90&-89.55\\
      78&86.63575&-53.75845&0.0000&   -&   -&       0&       0&19.95&21.23&23.73&0.69& -1.66\\
      79&86.63387&-53.75881&   -&   -&   -&       0&       0&22.33&24.91&25.17&0.29&-62.92\\
      80&86.65696&-53.75884&   -&   -&   -&       0&       1&22.34&24.42&  -&0.86& 50.68\\
      81&86.63404&-53.75795&   -&   -&   -&       0&       0&20.26&22.73&  -&0.53&-32.34\\
      82&86.64087&-53.75818&0.0000&   -&   -&       0&       0&20.14&21.48&24.05&0.75& 73.85\\
      83&86.65275&-53.75854&   -&   -&   -&       0&       1&21.91&24.12&24.86&0.79&-57.22\\
      84&86.65521&-53.75851&   -&   -&   -&       0&       0&22.65&25.05&  -&0.57& 10.22\\
      85&86.65500&-53.75820&1.0592&   -&   -&       1&       1&21.51&23.65&26.28&0.34& 39.16\\
      86&86.65408&-53.75843&   -&   -&   -&       0&       1&22.61&24.19&26.39&0.78& 44.23\\
      88&86.63317&-53.75823&0.3916&   -&   -&       0&       0&22.32&22.98&23.08&0.53&-26.28\\
      89&86.65483&-53.75834&   -&   -&   -&       0&       1&23.02&24.76&27.02&0.27& 49.41\\
      90&86.65742&-53.75806&   -&   -&   -&       0&       1&23.07&25.00&28.00&0.85& 36.59\\
      91&86.64396&-53.75743&   -&   -&   -&       0&       0&21.40&23.96&24.30&0.66& -7.60\\
      92&86.65483&-53.75720&   -&1.0647&1.0647&       1&       1&20.05&22.97&24.53&0.94&-31.48\\
      93&86.65162&-53.75729&   -&   -&   -&       0&       1&21.10&23.53&24.08&0.43& 10.21\\
      94&86.65746&-53.75743&   -&   -&   -&       0&       0&20.98&  -&25.13&0.68& 61.63\\
      95&86.62533&-53.75709&   -&   -&   -&       0&       0&20.39&22.66&23.40&0.54&-87.95\\
      96&86.62700&-53.75754&   -&   -&   -&       0&       0&22.09&23.72&23.47&0.69&  0.08\\
      97&86.65275&-53.75734&   -&   -&   -&       0&       1&21.65&24.17&25.24&0.86&-31.94\\
      98&86.65125&-53.75745&   -&   -&   -&       0&       0&23.63&26.75&27.06&0.41& 30.15\\
      99&86.64542&-53.75729&   -&   -&   -&       0&       0&22.14&24.74&24.75&0.67& 24.64\\
     100&86.65067&-53.75723&   -&   -&   -&       0&       0&22.60&  -&26.18&0.61& 13.73\\
     101&86.64158&-53.75720&   -&   -&   -&       0&       0&21.52&24.88&25.42&0.81&-72.49\\
     102&86.63850&-53.75676&   -&   -&   -&       0&       0&21.09&23.09&  -&0.43& 44.05\\
     103&86.63679&-53.75676&1.0783&   -&   -&       1&       1&20.56&23.34&23.66&0.84& 43.05\\
     104&86.62417&-53.75684&   -&   -&   -&       0&       1&22.02&24.29&24.27&0.94& 46.87\\
     105&86.65300&-53.75679&   -&   -&   -&       0&       0&22.05&  -&  -&0.57& 59.04\\
     106&86.65150&-53.75604&0.0000&   -&   -&       0&       0&20.61&22.21&23.31&0.13& 40.47\\
     107&86.64717&-53.75568&   -&   -&   -&       0&       1&20.41&22.92&23.66&0.44& 34.16\\
     108&86.63804&-53.75543&0.0000&   -&   -&       0&       0&18.24&19.87&20.59&0.61& 19.57\\
     109&86.64625&-53.75573&   -&   -&   -&       0&       0&22.72&24.78&25.23&0.58& 13.47\\
     110&86.62558&-53.75568&   -&   -&   -&       0&       1&21.21&23.66&22.57&0.66& 20.81\\
     111&86.65850&-53.75559&   -&   -&   -&       1&       1&22.26&24.35&24.88&0.66&-70.51\\
     112&86.65750&-53.75556&   -&   -&   -&       0&       0&22.05&26.16&35.59&0.69& 35.61\\
     113&86.64729&-53.75515&0.0000&   -&   -&       0&       0&21.54&22.59&  -&0.13& 35.15\\
     114&86.63567&-53.75509&   -&   -&   -&       0&       0&23.12&  -&24.10&0.81& 19.55\\
     115&86.64733&-53.75340&0.0000&   -&   -&       0&       0&16.45&19.52&20.64&0.87& 21.88\\
     116&86.65275&-53.75434&   -&   -&   -&       0&       1&21.33&24.15&25.15&0.39& 81.14\\
     117&86.65058&-53.75426&   -&   -&   -&       1&       1&21.84&24.35&25.43&0.83& 33.76\\
     118&86.64479&-53.75404&   -&   -&   -&       0&       0&22.35&  -&25.51&0.43& 75.26\\
     119&86.62475&-53.75384&0.0000&   -&   -&       0&       0&21.34&22.39&22.52&0.27&-28.27\\
     120&86.63892&-53.75379&   -&   -&   -&       0&       1&23.10&24.83&25.19&0.79& 40.62\\
     121&86.65237&-53.75315&   -&   -&   -&       0&       1&21.28&23.78&25.40&0.40& 54.29\\
     122&86.65162&-53.75268&   -&   -&   -&       0&       0&20.43&24.21&23.75&0.27& 67.01\\
     123&86.64612&-53.75168&   -&   -&   -&       0&       0&21.01&24.04&25.03&0.51&-26.23\\
     124&86.64087&-53.75168&   -&   -&   -&       0&       0&21.03&  -&22.88&0.47&-85.32\\
     125&86.62729&-53.75156&   -&   -&   -&       0&       1&21.79&23.87&24.54&0.67&-12.88\\
     126&86.64612&-53.75059&   -&1.0676&1.0676&       1&       1&19.60&22.75&24.04&0.70& 20.71\\
     127&86.64408&-53.74954&   -&   -&   -&       0&       0&21.95&25.46&25.20&0.64& 20.03\\
     128&86.65237&-53.74687&0.0000&   -&   -&       0&       0&19.70&21.06&22.90&0.70& 30.80\\
     129&86.65250&-53.74687&0.0000&   -&   -&       0&       0&19.93&21.30&22.51&0.49& 10.16\\
\hline
\end{longtable}
\end{centering}
{Columns: (1) Source ID; (2) right ascension; (3) declination; (4) redshift from GMOS spectroscopy (this work); (5) redshift from Brodwin et al. 2010; (6) redshift from Ruel et al. 2014; (7) 1 denotes spectroscopically-confirmed cluster member; (8) 1 denotes photometrically-classified cluster member; (9) $K_s$-band apparent magnitude; (10) F814W apparent magnitude; (11) F606W apparent magnitude; (12) {$K_s$-band} ratio of semi-major to semi-minor axis; (13) {$K_s$-band} position angle. Notes: Sources 69 \& 72 and 85 \& 89 are blended in our GMOS spectroscopy; see text for details.}
\twocolumn

\addcontentsline{toc}{chapter}{Bibliography}
\bibliographystyle{hapj}
\bibliography{GeMS_DR}

\appendix

\section{Distortion correction}
\label{distortion}

We include below instructions for calculating a customised distortion correction. This method is particularly useful when there are insufficient bright sources with known astrometry in the field of view; if there are sufficient bright sources in the field then the procedure can be commenced at (ii).
 SExtractor, SCAMP and SWarp are available from  {www.astromatic.net} \citep{Bertin1996,Bertin2002,Bertin2006}, GAIA \citep{Draper2014} from  {http://star-www.dur.ac.uk/$\sim$pdraper/gaia/gaia.html} and THELI \citep{Erben2005,Schirmer2013} from \href{https://astro.uni-bonn.de/~theli/}{https://astro.uni-bonn.de/$\sim$theli/}.

\begin{enumerate}
\item Observe a set of high signal-to-noise pointings of a star field, for which an excellent astrometry catalogue is available. We suggest the LMC or NGC288 depending on which is observable during your run. Use the same photometric band, position angle and similar airmass as science image, with similar guide star locations and relative magnitudes as far as possible. Resolution of the catalogue should be similar to observations, and magnitude range overlapping so that sufficient ($>$ tens of) stars are available per CCD. Format conversion can be done with THELI if necessary.
\item Calculate coarse correction to astrometric field headers using GAIA. This allows interactive input to the gross distortion calculation, and is particularly necessary when the image headers are significantly translated with respect to the true right ascension and declination.
\item Use the GAIA correction (header keywords CRVALi, CRPIXi, CDi\_j) to make an input .ahead file for SCAMP.
\item Calculate fine correction using SCAMP.
\item Apply SCAMP correction (except CRVAL) using SWarp.
SWarp must be run on each frame separately, as the telescope pointing information is not sufficient.
\item Coadd science frames based on the relative location of a bright star in the image.
\end{enumerate}

\section{Sky subtraction residual correction}
\label{apresidual}

When observing faint sources it is often desirable to use offset object frames as sky frames, in order to maximise the total time on source. However, faint sources are not detected in individual images so are not masked out from the offset frames, leading to residuals after sky subtraction. Here we outline our method for creating improved masks. This is particularly beneficial for observing with a semi-regular dither pattern, which may be desirable for extended sources or ensuring adequate coverage of the chip gaps.

\begin{enumerate}
\item Calculate the inverse correction on the astrometric field for each CCD, firstly with GAIA to make a coarse correction, then with SCAMP to calculate the fine correction.
\item Rotate the final stacked image and flip to the observed N is up, E is right reference frame. SCAMP and SWarp know about the true WCS, so cannot deal with this in the inverse correction.
\item Offset the stacked image by recalculating the reference pixels (CRPIXi) for each CCD in each observed image.
\item Apply the inverse correction with SWarp.
\item Cut out the region of the corrected image that matches the each observed CCD region.
\item Create a mask for each cutout.
\item Combine the masks with the {data quality (DQ)} extensions of the observed images, and proceed to the Gemini IRAF sky creation routine gasky.
\end{enumerate}

\section{PSF characterisation}
\label{psfchar}

Conventional AO imaging, which uses a single natural or laser guide star, results in a PSF that is composed of a diffraction-limited core profile plus a seeing-limited (typically Gaussian) uncorrected halo term, with minor smoothing due to tip-tilt jitter. The relative contribution of each term to the composite observed PSF is characterised by the Strehl ratio achieved during observation.
 
\begin{figure}
\centerline{
\includegraphics[width=1\linewidth]{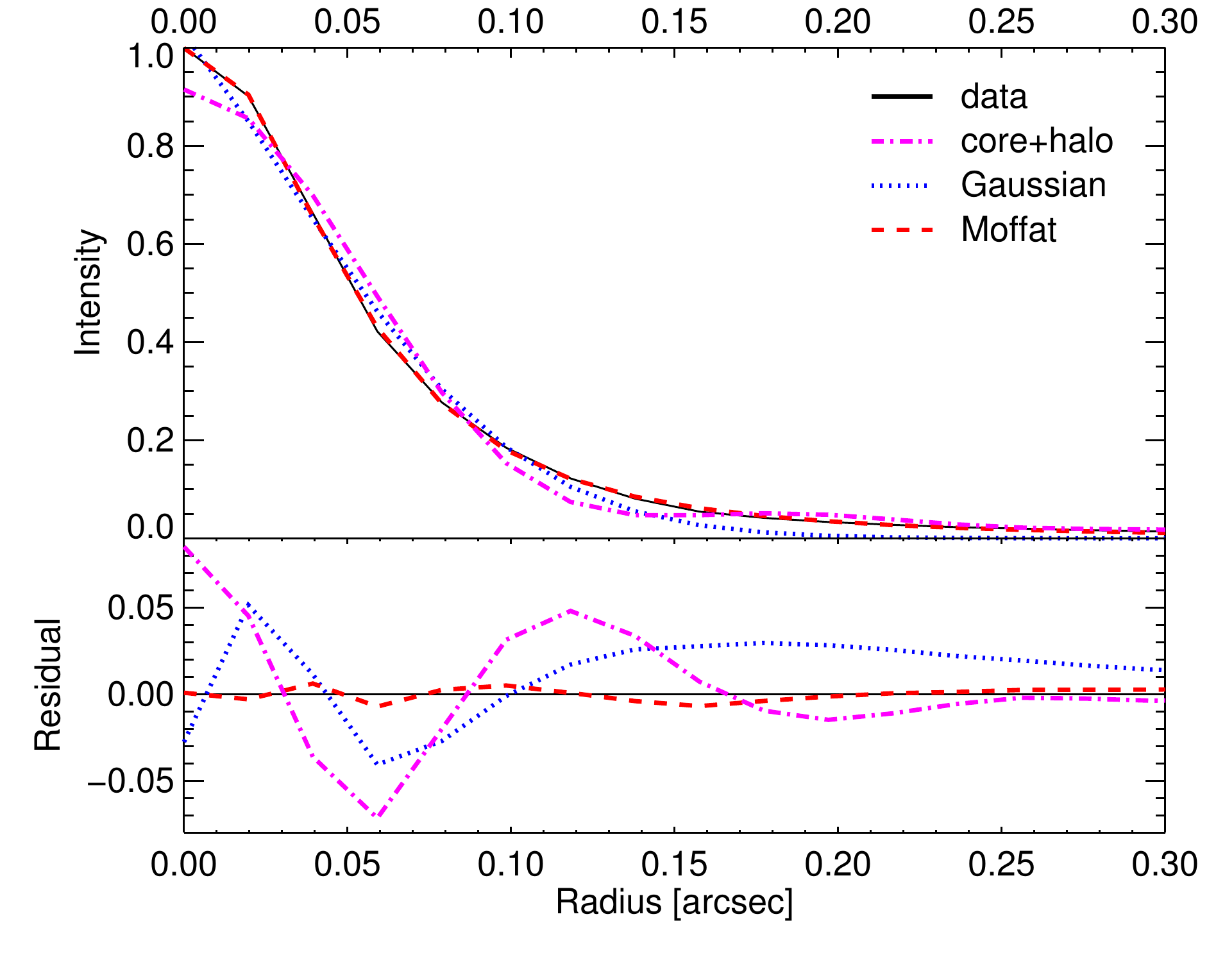}
}
\caption{\label{oned} {\it Top:} 1D star profile, fit with Airy core + Gaussian halo (magenta, dash-dot line), Gaussian (blue, dotted line) and Moffat (red, dashed line) profiles. Discontinuities in the data and fits are illustrative of the pixel sampling. {\it Bottom:} Residual profile. The observed profile is not well fit by the traditional AO-style core + halo shape, for the reasons explained in the text; the Moffat profile best represents the data.
}
\end{figure}

The nature of the MCAO imaging data presented in this work results in a PSF that is complex and varying. First, the laser asterism allows estimation of the high-order wavefront aberrations across GeMS, with the degree of correction degrading away from the field centre as error terms in the atmospheric reconstruction model grow. Next, due to the passage of the lasers' light up through the atmosphere prior to generation of the laser asterism in the sodium layer, natural guide stars are required to make the low-order (e.g., tip-tilt) correction. 
GeMS is equipped with three natural guide star probes, designed to observe a widely-separated three-star asterism spanning the GSAOI field. When three stars are not available, MCAO observations are still possible using two stars or even a single natural guide star, with the caveat that the quality of AO correction degrades with angular distance from the region nearest the star/s as the unprobed isokinetic error grows \citep{Rigaut2010}. Finally, an additional modification to the classic narrow-field AO core+halo PSF model is also appropriate due to the long composite exposure for our wide-field image. The total exposure time for our final stacked image is 4 hours 42 min, spread over three nights of observation at a range of airmasses and with variable natural seeing and laser guide star return power. The effect of the variable conditions over this extended observation period is to somewhat blur the diffraction-limited image core. The resulting PSF is illustrated in Figure~\ref{oned}. It is clear that the traditional AO profile does not fit the MCAO observations well.

The corollary of these observational necessities is that the PSF unavoidably varies across the GSAOI field of view. For dense star fields this poses little problem, since a high star density allows for excellent PSF modelling as a function of image location. For an extragalactic survey field (located at high Galactic latitude, with consequently few stars) it is much more challenging to quantify the variation in the PSF. For instance, we detect only eleven unsaturated stars in our imaging (including a close visual binary pair with separation of order the PSF). These were detected using a by-eye classification: eight brighter stars {($m_{F814W} < 22 $ AB mag)} have obvious diffraction spikes in the HST imaging, and a further three rather faint stars {($m_{F814W} \sim 22.5$ AB mag)}  have similar colours and SExtractor morphologies {to the brighter stars}. 

Ideally, target observations would be interspersed with regular observations of a sufficiently dense star field so as to provide a high quality model of the spatial variations in the PSF. However, since the PSF variation changes with telescope orientation and guide star asterism, such observations are only useful if the star field and science field are observed at the same orientation and with a similar asterism. Even when an appropriate star field can be identified, periodic observation of such a field would place an exceedingly high overhead on deep observations such as those presented in this work. An alternative methodology would be to develop a PSF simulation akin to the Tiny Tim \citep{Krist2011} simulator developed for characterisation of $HST$ PSFs. However, such an endeavour is not possible without access to sufficient reference data\footnote{A basic Fourier analysis of the imaging indicates that the residual wavefront error varies coherently, with the principal variation across the field being a trefoil aberration. However, due to the limited number of stars in the cluster field, and the associated signal-to-noise ratio of each star such an analysis was deemed unlikely to deliver significant improvement over the model adopted in this work.}.
 
\begin{figure}
\centerline{
\includegraphics[width=1\linewidth]{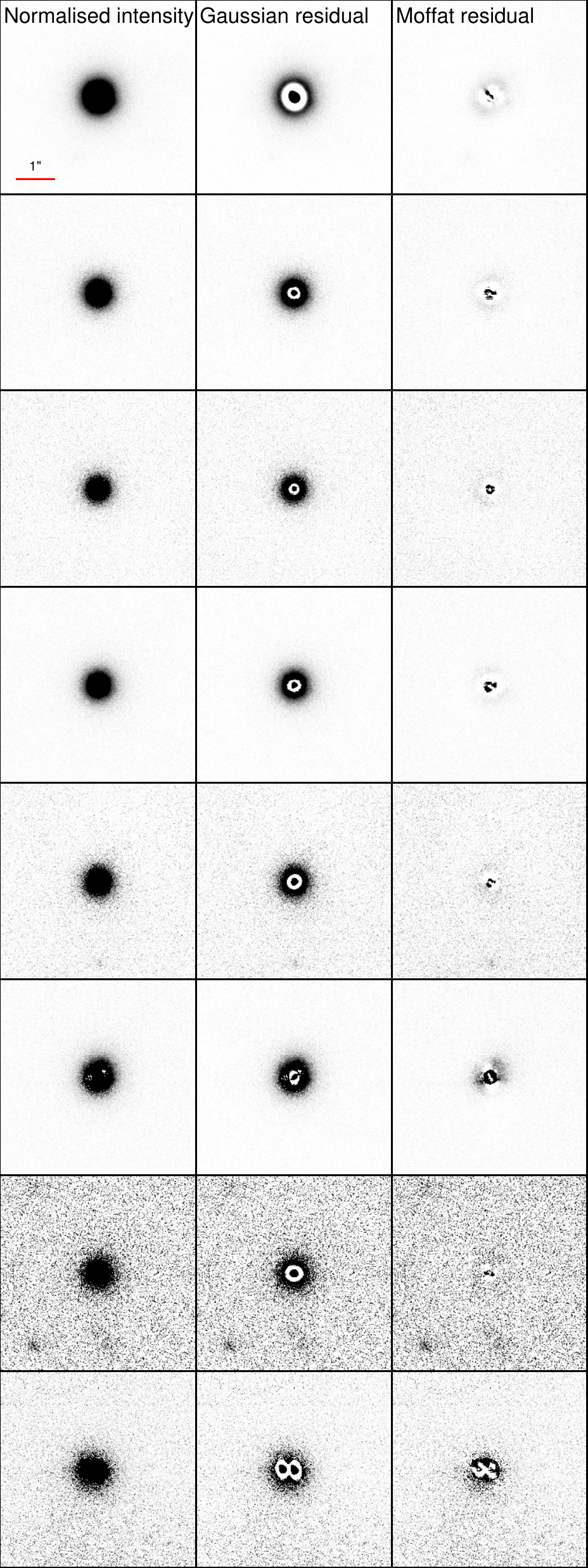}
}
\caption{\label{psffits} Profile fitting to known stars in the final GSAOI image, one star per row. Columns (left to right) show observed star, and Gaussian and Moffat residuals. Stars are normalised to the same intensity. The red scale bar is one arcsecond in length. The star in the bottom row is a binary, fitted by two function profiles with identical parameters except for amplitude and position.
%
}
\end{figure}

\begin{figure}
\centerline{
\includegraphics[width=1\linewidth]{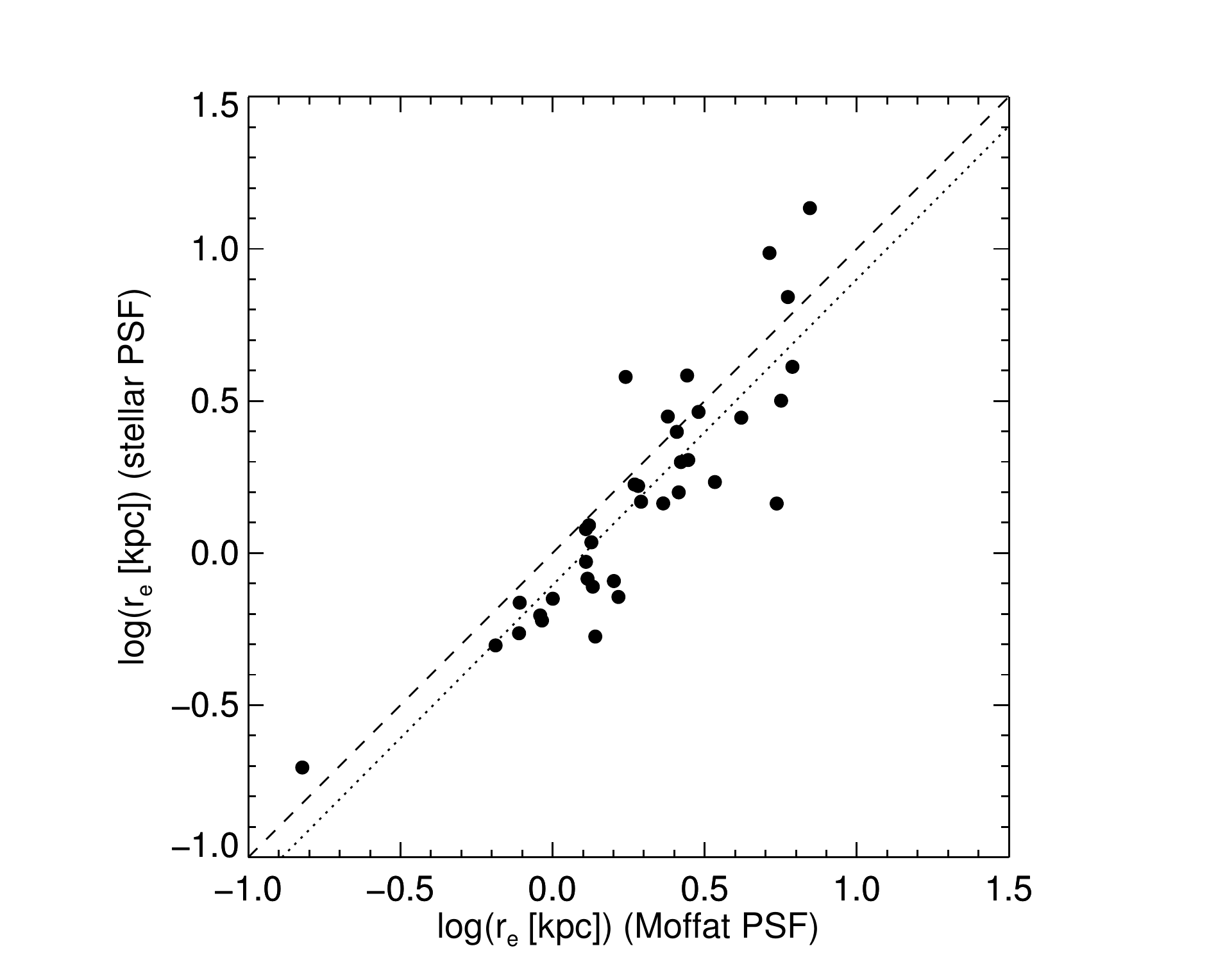}
}
\caption{\label{psffits1} Cluster member galaxy effective radii measured with two different PSFs: the x axis is constructed from an interpolated Moffat profile as used in this work; the y axis an empirical PSF of the nearest star to each galaxy. The dashed line shows the 1:1 relation, while the dotted line is the best fitting line.
}
\end{figure}

\begin{figure}
\centerline{
\includegraphics[width=1\linewidth]{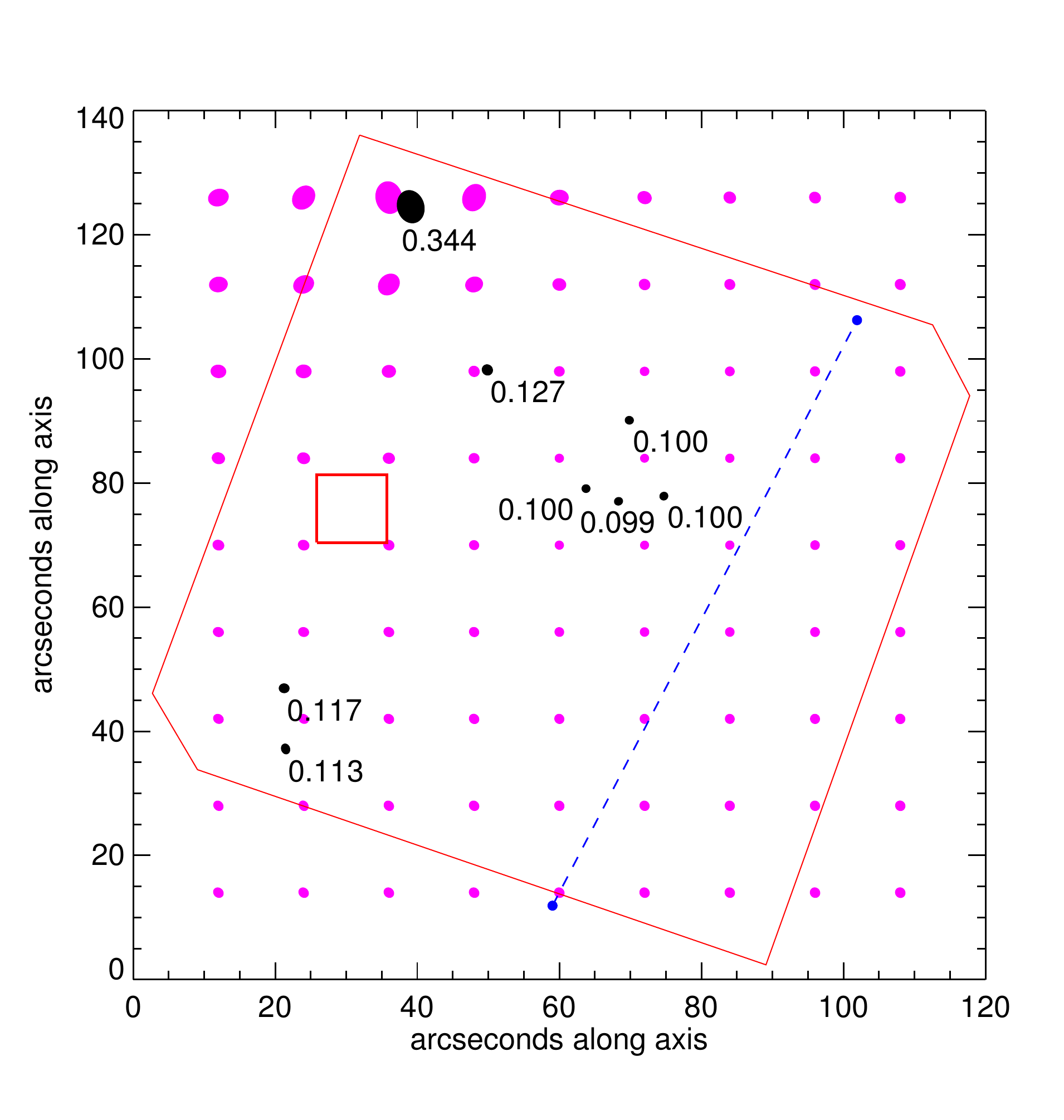}
}
\caption{\label{psf_plot} Variation of PSF FWHM and ellipticity across final stacked image, after discarding frames with mean FWHM $>$ 120 mas. Black circles indicate measured size and ellipticity of known stars, labelled with Moffat profile fit FWHM in arcseconds. Magenta circles indicate size and ellipticity interpolated across the field. The large, red polygon shows the sky coverage of our GSAOI image; the smaller, red square shows the cluster core location as in the inset of Fig.~\ref{sci}. Blue circles show the location of the natural guide stars, which are not measured in our imaging as one is saturated and the other does not fall within the field of view. The blue, dashed connecting line shows the region of theoretical best Strehl ratio (10.6\%); the ratio is predicted to degrade perpendicularly to this line. There are no cluster members near the star at the top of the field. 
}
\end{figure}

We chose the pragmatic compromise of fitting a low order parametric model to the eight brighter unsaturated stars in the GSAOI field. The stellar images are shown in the left-hand column of Figure~\ref{psffits}. Visual inspection indicates that a 2D PSF model of moderate ellipticity is required to fit these, with rotation of the major axis across the field. We show the residuals of a 2D elliptical Gaussian profile in the centre column of Figure~\ref{psffits}. Unsurprisingly, the Gaussian profile is inadequate as it has insufficient independent components to trace the complex modified MCAO profile described above. 
Ultimately, we adopt a 2D \citet{Moffat1969} profile as the simplest profile model capable of adequately representing the stellar PSF structure without introducing significant instability to the fitting procedure.
Fitting residuals for the identified stars with the Moffat PSF are shown in the right-hand column of Figure~\ref{psffits}. These are found to be at a level of $\sim$10\% of the stellar profiles. 
{Nevertheless, in Figure~\ref{psffits1} we demonstrate that effective radii measured with a modelled Moffat profile PSF are essentially the same as those measured with an empirical stellar PSF.}
We illustrate the variation in PSF parameters across the field in Figure~\ref{psf_plot}. This variation motivates us to construct a model PSF at the location of every source to be measured, rather than using an empirical profile measured from one or more of the stars within the field. The spatial variation is modelled by interpolating the best-fitting Moffat profile model parameters for each known star using an inverse distance interpolation. Note that this choice of a Moffat profile for the PSF is vindicated by \citet{Neichel2014a}, who found that the best functional form fitting the GeMS PSF is $(1+\alpha r^{2.4})^{-1}$, which is close to a Moffat profile with $\beta = 1.2$. This PSF shape is a natural consequence of the fact that the MCAO error budget is dominated --at least in the current incarnation of GeMS-- by the residual turbulence caused by turbulent layers in-between the 0- and 9-km deformable mirrors; this turbulence contributes a large, generalized fitting term.

\end{document}